\documentclass[sigconf, nonacm]{acmart}

\usepackage[linesnumbered,ruled,vlined]{algorithm2e}
\SetKw{KwOr}{or}
\SetKw{KwAnd}{and}
\usepackage{amsmath, amssymb,amsfonts}
\usepackage{graphicx}
\usepackage{textcomp}
\usepackage{natbib}
\usepackage{xcolor}
\usepackage{url}
\usepackage{booktabs}
\usepackage{multirow}
\usepackage{subfigure}
\usepackage{array}

\usepackage{pifont}  
\usepackage{balance}
\usepackage{textcomp}
\usepackage{comment} 
\usepackage{float}

\usepackage{rotating}
\usepackage{wasysym}
\usepackage{hyperref}
\usepackage{subcaption}
\usepackage{afterpage}
\usepackage{enumitem}
\usepackage{stfloats}
\usepackage{comment} 
\usepackage{rotating}
\usepackage{appendix}
\usepackage{placeins}
\usepackage[utf8]{inputenc}
\usepackage{enumitem}
\usepackage{microtype}
\setlist{nosep}
\usepackage{mathtools}
\usepackage{bm}
\usepackage{amsthm}
\usepackage{cleveref}
\usepackage{booktabs}
\usepackage{tabularx}
\usepackage{array}
\usepackage{xcolor}
\usepackage{tcolorbox}
\usepackage{fancyhdr}

\usepackage{microtype}
\usepackage[ruled,vlined,linesnumbered]{algorithm2e}
\begin{document}

\title{When Agents Talk, Secrets Travel: Topology-Driven Leakage and Recipient-Conditioned Sanitization in Multi-Agent LLM Systems}

\author{Sunday Oyinlola Ogundoyin}
\email{sunday.ogundoyin@hdr.mq.edu.au}
%\orcid{0000-0003-1564-2818}
\affiliation{%
  \institution{Macquarie University}
  \city{Sydney}
  \state{NSW}
  \country{Australia}
}

\author{Muhammad Ikram}
%\orcid{0000-0003-2113-3390}
\affiliation{%
  \institution{Macquarie University}
  \city{Sydney}
  \state{NSW}
  \country{Australia}}
\email{muhammad.ikram@mq.edu.au}

% \author{Rahat Masood}
% \orcid{0000-0001-5935-0062}
% \affiliation{%
%   \institution{The University of New South Wales}
%   \city{Sydney}
%   \state{NSW}
%   \country{Australia}
% }

% \author{Sunday Oyinlola Ogundoyin}
% \affiliation{%
%   \institution{Macquarie University}
%   \city{Sydney}
%   \state{NSW}
%   \country{Australia}
% }
% \email{sunday.ogundoyin@hdr.mq.edu.au}

% \author{Muhammad Ikram}
% \affiliation{%
%   \institution{Macquarie University}
%   \city{Sydney}
%   \state{NSW}
%   \country{Australia}
% }
% \email{muhammad.ikram@mq.edu.au}

% \author{Hassan Jameel Asghar}
% \affiliation{%
%   \institution{Macquarie University}
%   \city{Sydney}
%   \state{NSW}
%   \country{Australia}
% }
% \email{hassan.asghar@mq.edu.au}

% \author{Benjamin Zi Hao Zhao}
% \affiliation{%
%   \institution{Macquarie University}
%   \city{Sydney}
%   \state{NSW}
%   \country{Australia}
% }
% \email{ben_zi.zhao@mq.edu.au}

% \author{Mohamed Ali Kaafar}
% \affiliation{%
%   \institution{Macquarie University}
%   \city{Sydney}
%   \state{NSW}
%   \country{Australia}
% }
% \email{dali.kaafar@mq.edu.au}
\renewcommand{\shortauthors}{Sunday et al.}
\sloppy
    
\begin{abstract}

Multi-agent LLM systems (MALS) introduce a privacy-leakage problem that conventional sensitivity-based defenses miss, since information disclosed to one authorized agent can propagate through routine coordination to an unauthorized one, without compromise or malicious behavior. This motivates topology-aware enforcement that blocks unauthorized propagation while preserving authorized disclosure. We evaluate 485 scenarios across coordinator-worker, peer-to-peer, and hierarchical topologies, instrumenting seven communication channels and measuring surface leakage and disclosure to topology-introduced recipients. We introduce RECONS, a \emph{Detect $\rightarrow$ Decide $\rightarrow$ Transform} sanitizer conditioning intervention on recipient and execution context rather than sensitivity alone. Topology changes where protected information propagates, with disclosure to topology-introduced recipients rising from 0\% to 52.2\% while surface leakage stays nearly flat at 60.6–62.8\%. This effect persists across four generator models and a second orchestration framework, though magnitude and relay paths differ; framework choice separately drives a 150× leakage spread, so topology alone is not sufficient. RECONS reduces residual disclosure from 96.7\% without sanitization to 29.25\%, preserving 94.1\% of authorized disclosures versus 20.5\% under blanket redaction. Detection coverage is the dominant bottleneck, accounting for 95.5\% of residual disclosures. These results show that privacy enforcement in MALS must reason about the recipient and communication path realized at runtime, not sensitivity or topology assumptions alone.

\end{abstract}

% \keywords{Clinical hallucination, medical GPTs, privacy policy, OpenAI, LLM-as-a-judge}

\maketitle

%%%%%%%%%%%%%%%%%%%%%%%%%%%%%%%%

\section{Introduction}\label{sec:introduction}

Large Language Model (LLM) applications are evolving from single-agent interactions into multi-agent systems in which specialized agents collaborate to complete task~\cite{tran2025multiagentcollaborationmechanismssurvey, Zhang_collaboration}. One agent may plan a task, another may retrieve information or invoke a tool, and others may coordinate, maintain state, or produce intermediate artifacts. This decomposition improves modularity and capability, but it also changes the security model for LLM applications. Information is no longer exchanged only between a user and one model; it can propagate across multiple agents and execution channels before the system produces its final response. This turns collaboration in multi-agent LLM systems (MALS) into an information-flow problem~\cite{asif, costa2025}, in which sensitive-data leakage occurs not because a value is disclosed, but because it reaches the wrong recipient. Consider a user $u$ discloses a protected value $v$ to agent $A$, who is authorized to receive it; coordinating with $B$ to complete the task, $A$ forwards $v$ to $B$, who is not authorized to receive it. No agent is compromised, and no malicious prompt is required. The violation arises not from the initial disclosure but from its propagation past the authorization boundary.

This differs from prior treatments of multi-agent leakage in three
ways. First, exposure changes with topology. Adding an agent adds a
new party the value can reach, and a value confined to a two-agent
exchange may be forwarded to an agent that exists only because the
system grew richer. AgentLeak~\cite{AgentLeak} measures
internal-channel leakage under one fixed two-agent topology and does not vary it. Liu et al.~\cite{liu-etal-2026-topology-matters} show topology affects leakage magnitude, but under an adversary actively probing memory, not benign execution with an authorization relation over recipients. Neither establishes whether a legitimate disclosure becomes unauthorized solely through coordination structure. Second, a defense treating every detected value alike cannot distinguish these cases; it must sanitize everything, destroying legitimate disclosures, or sanitize nothing, permitting unauthorized ones. Third, most existing sanitizers do not decide whom a value may reach. Preempt~\cite{Preempt} transforms a value once judged sensitive, but for one user and one model owner, with no second, differently
authorized recipient to condition on. Reasoning-based mitigations~\cite{PrivacyChecker,li_1-2-3_nodate} improve how an
agent judges disclosure, but the judgment remains the model's own
about a single message, not a policy conditioned on where it is
forwarded. FIDES~\cite{costa2025} enforces provable guarantees over execution traces, but targets adversarial prompt injection, not benign over-disclosure. All treat sensitivity and authorization as one question---but they are two.

We refer to this as \emph{topology-conditioned recipient exposure}: a coordinator--worker (CW), a peer-to-peer (P2P), and a hierarchical (Hier) system expose different recipient sets, so changing topology changes where an already-disclosed value propagates even when task, value, and initial authorization are unchanged. Richer topology need not monotonically increase leakage; it changes the recipient set against which authorization is evaluated. This motivates measuring how information propagates across communication structures and enforcing privacy selectively at recipient boundaries. This paper answers three questions.

\begin{itemize}[leftmargin=*] 
\item \textbf{RQ1:} How does topology affect propagation to
unauthorized recipients, and does this hold across frameworks and
models?
\item \textbf{RQ2:} Does recipient-conditioned enforcement reduce
unauthorized disclosure while preserving authorized disclosure?
\item \textbf{RQ3:} What accounts for residual exposure after
sanitization?
\end{itemize}

For \textbf{RQ1}, we instantiate the same scenarios across CW, P2P,
and Hier topologies, measuring both disclosure and whether it
reaches an unauthorized recipient. Topology changes propagation and
exposes values to topology-introduced recipients, persisting across
models and frameworks, though magnitude and ordering depend on
execution semantics. Holding topology fixed, framework choice alone
produces a leakage spread of roughly two orders of magnitude;
topology is thus security-relevant but not a sufficient predictor on
its own.

RQ1 gives the second distinction a concrete mechanism: topology is
how a value authorized for one recipient reaches another it never
was, which motivates \textbf{RQ2}. If a detected value $v$ is
authorized for the recipient, transforming it destroys legitimate
information; if unauthorized, passing it through permits the
violation. Recipient-blind enforcement must choose: blanket
redaction over-suppresses, pass-through under-protects. The missing decision is not whether $v$ is sensitive, but whether \emph{this recipient may receive $v$ here}. We address this with \textbf{RECONS}, an inline, recipient-conditioned mechanism. A value should be transformed when about to cross an authorization boundary to an unauthorized recipient, not merely because it is sensitive. RECONS separates enforcement into \textsc{Detect} $\to$ \textsc{Decide} $\to$ \textsc{Transform}: detection identifies candidates, decision evaluates recipient authorization, and transformation intervenes only when disclosure is unauthorized, preserving authorized disclosure and falling back conservatively when resolution or safe transformation is unavailable.

Compared to no sanitization, blanket redaction, and recipient-blind DLP, RECONS reduces residual disclosure from 96.7\% to 29.25\% (28.3\% blanket, 35.5\% Presidio) while retaining 94.1\% of authorized disclosures, roughly $4.6\times$ blanket redaction's. This also answers \textbf{RQ3}: 95.5\% of residual cases are detection misses, a ceiling any same-detector sanitizer would share; 4.5\% are recipient-unresolvable, specific to RECONS's own resolution step; seven remain post-transformation detector-positive. Privacy enforcement in MALS must therefore account for where information is routed, not only what it is: topology determines reachable recipients, authorization determines permitted disclosures, and an effective control decides at the communication boundary.

This paper makes the following contributions:
\begin{enumerate}[leftmargin=*]
\item \textbf{Topology- and architecture-conditioned recipient-exposure characterization.} We characterize protected-value propagation across CW, P2P, and Hier topologies and replicate it across generator models and frameworks. We show that topology and framework choice independently drive recipient exposure, with framework alone producing a roughly two-order-of-magnitude spread.

\item \textbf{Recipient-conditioned privacy enforcement.} We introduce RECONS, an inline sanitizer implemented as a \textsc{Detect} $\rightarrow$ \textsc{Decide} $\rightarrow$ \textsc{Transform} pipeline that conditions enforcement on recipient identity and authorization context whenever observable.

\item \textbf{Controlled privacy--authorization evaluation.} We evaluate RECONS against no sanitization, blanket redaction, and recipient-blind DLP under a frozen evaluation protocol. RECONS approaches blanket redaction's residual-leakage performance while preserving substantially more authorization-permitted disclosures.

\item \textbf{Residual-exposure decomposition.} We distinguish detection misses, recipient-resolution failures, and post-transformation residual cases, identifying detection coverage, a limitation any downstream sanitizer would share, as the dominant bottleneck, and recipient resolution as a limitation specific to this architecture.
\end{enumerate}

\begin{figure*}
    \centering
    \small
\includegraphics[width=0.9\linewidth]{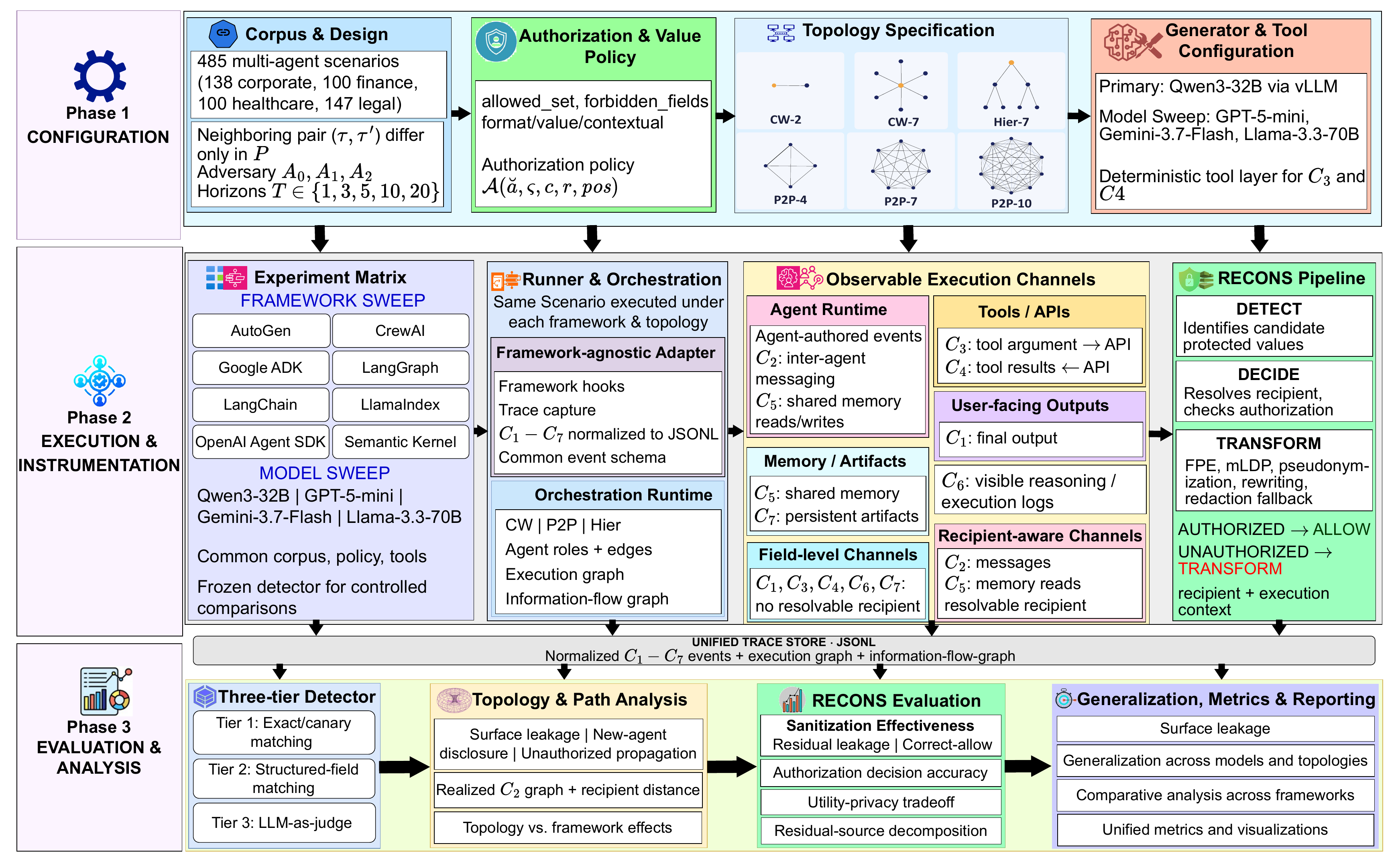}
\vspace{-0.4cm}
    \caption{\small Overview of the experimental pipeline: configuration (corpus, topology, and authorization policy), execution (framework and model sweeps), and evaluation (detection, RECONS sanitization, and cross-condition analysis).}
    \label{fig:overview}
\end{figure*}

\noindent \autoref{fig:overview} summarizes the resulting pipeline, from corpus and topology configuration through framework- and model-level execution to detection and RECONS-based evaluation.

\section{Background and Related Work}\label{sec:related-work}

\subsection{Background}\label{sec:background}

A multi-agent LLM system decomposes a task across multiple LLM-driven agents coordinated by a framework (e.g., LangChain~\cite{Langchain}, AutoGen~\cite{AutoGen}, LangGraph~\cite{Langgraph}) in a CW, P2P, or Hier arrangement; production frameworks typically support more than one. This decomposition means an agent's action is not confined to a single request-response exchange: it can invoke tools, read and write shared memory, and message other agents directly, each a point where information available to one agent becomes available elsewhere. So, a value disclosed once can reach further participants purely because of how the system is wired, independent of whether that disclosure was itself appropriate, the property our threat model (\S\ref{sec:threat-model}) formalizes as observable channels and a topology-dependent recipient set. 

Addressing this exposure requires three capabilities: detecting disclosure, deciding whether the resolved recipient is authorized, and transforming an unauthorized value without destroying its utility for authorized ones. Detection and transformation draw on existing techniques; deciding based on the recipient rather than sensitivity alone is the focus of our mechanism, introduced in
\S\ref{sec:decision-layer}. LLM-as-judge detection scores semantic equivalence between a candidate span and a protected value using a separate, calibrated model, catching paraphrase-based disclosure that pattern matching cannot, once its threshold is calibrated against labeled examples (\S\ref{sec:detection-layer}). Once a value is detected and unauthorized, format-preserving encryption (FPE) keeps its ciphertext in the plaintext's format, letting it pass through downstream logic built for that format, at the cost of a minimum domain size. Metric local differential privacy (mLDP) instead perturbs a value under a privacy budget $\varepsilon$ for approximate utility, while pseudonymization substitutes a deterministic, keyed value preserving referential consistency but no content. Preempt~\cite{Preempt} applies FPE and metric-LDP for a single user, and our transformation layer (\S\ref{sec:transformation-layer}) applies both per-channel, conditioned on the recipient.

\subsection{Related Work}
Privacy in MALS has been studied along three separate lines: whether agents disclose sensitive information, whether multi-agent execution creates new leakage, and how defenses should intervene once a value is detected, but none conditions its central decision on who the value is about to reach.

\subsubsection{Privacy Leakage in Multi-Agent LLM Systems.}\label{subsec:bg-leakage-phenomena}
Some work investigates whether an agent judges disclosure correctly. CONFAIDE~\cite{ConfAIDe} evaluates contextual-integrity reasoning directly and finds model-human agreement on appropriate information flow drops from 0.8 to 0.1 across four tiers of complexity. PrivacyChecker~\cite{PrivacyChecker} reports the same failure as a judgment-action gap, where models identify sensitive content correctly 98\% of the time yet leak it in roughly a third of interactions, and its live benchmark, PrivacyLens-Live, shows higher leakage than static single-turn evaluation. 1-2-3-Check~\cite{li_1-2-3_nodate} reduces leakage by decomposing privacy reasoning into extraction, checking, and execution agents. All three treat disclosure as a decision the acting model makes about a single message; none conditions it on where the message propagates afterward.

Other work shows leakage is not confined to what one agent discloses. Patil et al.~\cite{patil2025} formalize compositional leakage: an adversary aggregating benign responses from multiple agents reconstructs attributes no single agent stated,
and show chain-of-thought prompting blocks only 31--39\% of such reconstructions. Asif and Amiri~\cite{asif} derive an information-theoretic bound with the same result. Biswas et al.~\cite{Biswas_quantifying_privacy_leakage} quantify leakage with a combined mutual-information, Bayesian-inference-gain, and R\'enyi-DP framework. They showed that leakage increases with topology depth and collusion. MAGPIE~\cite{juneja2025_magpie} finds leakage up to 56\% in negotiation even under explicit no-disclosure instructions. GAMA~\cite{GAMA} and SRPG~\cite{SRPG} respond architecturally, isolating raw data in a private workspace or reconstructing sanitized content. However, none ties leakage magnitude to communication-graph structure or an explicit recipient-authorization relation, the two properties central to our threat model.

\subsubsection{Multi-Agent Leakage Measurement.}\label{subsec:bg-measurement}

AgentLeak~\cite{AgentLeak} shows inter-agent messages and shared memory leak the protected attribute at $2.1\times$ the external output's rate, output-only auditing misses 41.7\% of violations, and blanket redaction drops task success by 30--65 points. However, its evaluation uses one fixed two-agent topology. The authors established that internal channels leak, but not how communication structure affects where a disclosed value subsequently reaches. 

Liu et al. present MAMA~\cite{liu-etal-2026-topology-matters},
showing that denser, more centralized topologies and shorter attacker-target distance increase PII memory leakage across six graphs, the closest prior evidence that topology affects multi-agent privacy. Two differences remain: MAMA's threat model actively seeds and extracts memory over repeated probing, while we measure organic propagation during benign execution; and MAMA reports leakage as a function of graph distance and centrality, while we define it relative to an authorization relation over (value, recipient, channel) triples. So the question is not whether a value reaches a farther node but whether that node was authorized to receive it. Biswas et al.'s topology-depth finding~\cite{Biswas_quantifying_privacy_leakage} is consistent from an information-theoretic angle but, like MAMA, is not tied to an authorization boundary or a recipient-conditioned defense. AgentBench~\cite{liu2025agentbenchevaluatingllmsagents} and AgentDojo~\cite{AgentDojo} evaluate task competence and prompt-injection robustness, not internal-channel exposure.
OPENAGENTSAFETY~\cite{vijayvargiya_openagentsafety_2025} evaluates privacy alongside other harms in a static benchmark, not a topology-conditioned measurement. To our knowledge, no prior measurement work isolates topology as an independent variable while holding scenario, detector, and policy fixed, as our RQ1 requires.

\subsubsection{DLP and Privacy-Preserving Transformation.}\label{subsec:bg-dlp}

Pr$\in\in$mpt~\cite{Preempt} sanitizes a single prompt with FPE for format-dependent tokens and mLDP for value-dependent ones, with a formal indistinguishability guarantee, but for one outbound prompt and its response, with no second, differently authorized recipient.
RECONS reuses these mechanisms per channel but adds a recipient-aware decision layer that Pr$\in\in$mpt's single-agent setting does not require: not only how to transform a protected value, but whether a given recipient requires it transformed at all. 
Multi-Agent Firewall~\cite{cuesta2026-multi-agent-firewall-architecture-privacy}
intercepts outbound traffic with deterministic and NER-based detectors before a third-party provider, while AgentGuard~\cite{agentguard} enforces attribute-based access control on tool calls; both decide whether content crosses a perimeter, not how a value already inside a multi-agent system should be treated differently by an internal recipient. Our sanitizer targets that internal gap, motivated by AgentLeak's finding that internal channels leak at $2.1\times$ the external rate.

\subsubsection{Context- and Recipient-Aware Enforcement.}\label{subsec:bg-context-aware}

FIDES~\cite{costa2025} enforces noninterference via dynamic label propagation, blocking all policy violations on AgentDojo~\cite{AgentDojo} with fewer task losses than naive taint tracking, but targets an adversarial threat (indirect prompt injection). We study benign disclosure, defining authorization over observable triples rather than label lattices, trading formal guarantees for applicability to black-box traces. Sentinel~\cite{Sentinel} monitors a shared conversational space and quarantines violations, but is only a proof-of-concept study without an authorization taxonomy or utility trade-off. PrivacyChecker's sender, recipient, subject, and transmission principle is the closest reasoning-based analog. Still, judgment is made by prompting the acting model itself, inheriting the judgment-action gap it was meant to diagnose rather than using a frozen decision layer. Ling et al.~\cite{ling2026} find that persistent-state and propagation threats remain underserved relative to prompt-injection and tool-hijacking, the imbalance this paper corrects.\\

Table~\ref{tab:related-work-comparison} summarizes the comparison.
AgentLeak shows internal channels leak and redaction is costly under
one fixed topology; MAMA and Biswas et al. show topology affects
leakage magnitude under adversarial extraction, without recipient authorization. PrivacyChecker, 1-2-3-Check, and FIDES condition enforcement on some context, sender or recipient judgment, decomposed reasoning, or label flow, but none on the topology-dependent recipient for benign disclosures. Preempt and the DLP systems in \S\ref{subsec:bg-dlp} provide transformation and perimeter enforcement, not internal propagation control. This paper
extends topology-conditioned propagation beyond AgentLeak's single topology and adds the authorization-aware decision layer that single-agent and perimeter systems lack.

\begin{table}[t]
\centering
\tabcolsep = 0.07cm
\small
\caption{\small Comparison with the closest prior measurement and defense
work. A benign threat model measures organic propagation; propagation control means the mechanism acts on where a value is headed, not only whether it is sensitive.}
\vspace{-0.4cm}
\label{tab:related-work-comparison}
\begin{tabular}{@{}lcccc@{}}
\toprule
& Topology & Recipient- & Benign threat & Propagation \\
& varied & conditioned & model & control \\
\midrule
AgentLeak~\cite{AgentLeak}      & \ding{55} & \ding{55} & \ding{51} & \ding{55} \\
MAMA~\cite{liu-etal-2026-topology-matters} & \ding{51} & \ding{55} & \ding{55} & \ding{55} \\
Biswas et al.~\cite{Biswas_quantifying_privacy_leakage} & \ding{51} & \ding{55} & \ding{55} & \ding{55} \\
PrivacyChecker~\cite{PrivacyChecker} & \ding{55} & \ding{55}$^\dagger$ & \ding{51} & \ding{55} \\
1-2-3-Check~\cite{li_1-2-3_nodate}   & \ding{55} & \ding{55} & \ding{51} & \ding{55} \\
FIDES~\cite{costa2025}          & \ding{55} & \ding{55} & \ding{55} & \ding{51}$^\ddagger$ \\
Preempt~\cite{Preempt}          & \ding{55} & \ding{55} & \ding{51} & \ding{55} \\
\textbf{This paper}             & \ding{51} & \ding{51} & \ding{51} & \ding{51} \\
\bottomrule
\end{tabular}

\vspace{4pt}
\parbox{\linewidth}{\footnotesize $^\dagger$Reasons over
sender/recipient/subject but enforces via the acting model's own
prompted judgment. $^\ddagger$Controls propagation via label noninterference under an adversarial prompt-injection threat model.}
\end{table}

\section{Threat Model and Problem Formulation}\label{sec:threat-model}

\subsection{Observable Channels and Multi-Agent Execution} \label{subsec:channels}

We investigate privacy leakage in MALS as it manifests on \emph{observable channels}, points where information leaves an agent's private context and becomes available to other entities. Following AgentLeak~\cite{AgentLeak}, we consider seven channels, complete for our threat model: C1, final output to the user; C2, inter-agent messaging; C3 and C4, arguments supplied to a tool call and the result returned; C5, shared memory reads and writes; C6, visible reasoning and execution logs; and C7, artifacts the system produces. These differ in how a protected value is represented, which determines what a detector can act on. C1, C2, and C6 are \emph{free-text}, where values appear embedded in prose, often paraphrased rather than quoted, since a delegation message retells its originating task in the model's own words. C3 and C5 are \emph{structured}, placing a value in a discrete typed field, while C4 and C7 are \emph{mixed}. Only C2 and C5 reads additionally carry
an addressee that the trace identifies; the rest expose a value without naming a recipient, which later bounds recipient-aware enforcement (\S\ref{sec:decision-layer}).

We represent a single observable event as a tuple $\vartheta_t = (\breve{a}, \varsigma, c, r, \mathit{pos}, t)$, where $\breve{a}$ is the protected attribute, $\varsigma$ its value type (\S\ref{subsec:value-taxonomy}), $c \in \{\text{C1}, \dots, \text{C7}\}$ the channel, $r$ the recipient when resolvable, $t$ the temporal index, and $\mathit{pos} \in \{\mathit{pos}_i,
\mathit{pos}_j\}$ records whether $r$ is one of the scenario's original, presumed-authorized agents or an agent the topology introduces (\S\ref{subsec:topology-model}).

\subsection{Authorization Semantics}\label{subsec:authorization}
Whether an occurrence of $\breve{a}$ on channel $c$ is a violation depends not only on sensitivity but on \emph{who} receives it. We formalize this as an authorization relation $\mathcal{A}(\breve{a}, \varsigma, c, r, \mathit{pos}) \in \{\textsc{allow}, \textsc{sanitize}\}$, an \emph{externally specified ground-truth relation} derived from each scenario's own role and
access-control definitions. RECONS's decision layer (\S\ref{sec:sanitizer}) is evaluated against $\mathcal{A}$ as an oracle; an occurrence with $\mathcal{A}(\cdot) = \textsc{sanitize}$ is an \emph{authorization-boundary crossing}.

Conditioning $\mathcal{A}$ on $r$ and $\mathit{pos}$, rather than $\breve{a}$ and $c$ alone, is the formal distinction from prior models: AgentLeak~\cite{AgentLeak} defines leakage independently of recipient identity, and Pr$\in\in$mpt~\cite{Preempt} assumes a single user and model owner, so neither captures recipient-specific authorization, a value authorized for one recipient is not thereby authorized for a different, topology-reachable one. RECONS enforces this directly: $\mathcal{A}(\breve{a},\varsigma,c,r,\mathit{pos}_j)=\textsc{sanitize}$. Possession by an authorized agent is not a violation; a violation occurs when topology exposes the value to a reachable recipient unauthorized for that field.

\subsection{From Boundary Crossing to Disclosure}\label{subsec:boundary}
The authorization function determines a crossing but does not itself imply disclosure, which we define as the subsequent observation of $\breve{a}$ in channel $c$'s content at an unauthorized recipient. Leakage requires both. Without RECONS (\S\ref{sec:rq1}), no transformation intervenes, so every crossing exposing $\breve{a}$ is realized as disclosure and counted as leakage. With RECONS (\S\ref{sec:sanitizer}), a crossing counts as leakage only when transformation fails to prevent recovery of $\breve{a}$ from $c$,
separating whether a recipient is authorized from whether the value is recoverable once received: an authorized recipient observing $\breve{a}$ is not a violation, an unauthorized one who cannot recover it after transformation is a crossing without leakage, and one who can recover it is leakage.

\subsection{Topology Model}\label{subsec:topology-model}
We model a topology as $T=(\mathrm{\breve{S}},k)$, where $\mathrm{\breve{S}}$ specifies the coordination pattern and $k$ the agent count, across six cells: star (CW-2, CW-7), P2P (P2P-4, P2P-7, P2P-10), and hierarchical (Hier-7). CW-2 is the reference cell: with only the original two agents, no $\mathit{pos}_j$ recipient can exist, so exposure is $0\%$ by construction. It sits on neither axis, anchors the design to the original scenarios and cross-framework baseline, and is the baseline the rest are evaluated against in \S\ref{sec:rq1}. The cells impose no total ordering over complexity; agent count does not monotonically increase exposure, and relative leakage among non-reference topologies depends on framework and generator (\S\ref{subsec:framework-generalization}--\S\ref{subsec:model-generalization}). Our claim is limited to the structural distinction: every non-reference topology admits $\mathit{pos}_j$ recipients, creating disclosure paths CW-2 lacks. Non-reference cells are constructed by re-roling the same scenarios without expanding authorization policies; we analyze effects through paired within-scenario comparisons.

The \textbf{structure axis} $\{\text{CW-7}, \text{P2P-7}, \text{Hier-7}\}$ holds $N=7$ and varies wiring: CW-7 is a star with one coordinator and six workers, P2P-7 is fully connected, and Hier-7 is a two-level hierarchy of one root, two sub-coordinators, and four workers. The \textbf{scale axis} $\{\text{P2P-4}, \text{P2P-7}, \text{P2P-10}\}$ holds peer-to-peer wiring fixed and varies $N \in \{4, 7, 10\}$; the axes share P2P-7 as both a structure comparison and a middle-scale point (\autoref{fig:overview}).

\subsection{Protected-Value Taxonomy}\label{subsec:value-taxonomy}
We classify protected values into three classes by applicable transformation (Table~\ref{tab:value-taxonomy}). Format-dependent values (identifiers, account numbers, dates of birth) are recognizable by structure and support format-preserving transformation. Value-dependent values (names, addresses, numeric
amounts) are sensitive by content and support pseudonymization or mLDP. Contextual values (diagnoses, legal strategies, internal assessments) derive sensitivity from meaning and require semantic rewriting or redaction. This extends Pr$\in\in$mpt's two-class distinction~\cite{Preempt} with a contextual class arising where sensitive information sits in free-form agent-to-agent text rather than a discrete field; the three classes determine which transformation applies (\S\ref{sec:sanitizer}).
\begin{table}[t]
\centering
\small
\tabcolsep = 0.07cm
\caption{\small Protected-value taxonomy, defined by admitted transformation.}
\vspace{-0.4cm}
\label{tab:value-taxonomy}
\begin{tabular}{@{}lll@{}}
\toprule
Value type $\tau$ & Sensitive by & Admitted transformation \\
\midrule
Format-dependent & recognizable format & format-preserving encryption \\
Value-dependent  & the information carried & pseudonymization / mLDP \\
Contextual       & meaning in context & semantic rewriting / redaction \\
\bottomrule
\end{tabular}
\end{table}
\subsection{Adversary Model}\label{subsec:adversary-model}
We model a \emph{passive unauthorized recipient} observing one or more instrumented channels $C_1$--$C_7$ without authorization for the value under $\mathcal{A}$, canonically a $\mathit{pos}=\mathit{pos}_j$ agent introduced by the topology; an observer of shared logs ($C_6$) or persistent artifacts ($C_7$) is also in scope when unauthorized for that field. The adversary does not modify the execution, inject content, or obtain information beyond channels the scenario exposes, and we assume no computational limit on interpreting observed content.

We consider three adversary levels~\cite{AgentLeak}: $A_0$, benign execution; $A_1$, an external adversary targeting $C_1$; and $A_2$, an internal adversary targeting $C_2$--$C_5$. The primary threat model excludes inference from indirect evidence, collusion among unauthorized recipients, cross-session correlation, and observation outside $C_1$--$C_7$; these stronger capabilities are evaluated separately in \S\ref{subsec:reconstruction-results}--\S\ref{sec:behavioral-results} and excluded from the primary leakage measurements in \S\ref{sec:rq1} and sanitizer evaluation in \S\ref{sec:sanitizer}.

\section{Experimental Methodology}\label{sec:methodology}
Our experiments vary topology, adversary level, protected-value class, observable channel, and execution horizon. Unless stated otherwise, the generator and framework are fixed to Qwen3-32B and LangGraph, isolating topology as the independent variable.

\subsection{Corpus Construction and Evaluation Scale}\label{subsec:corpus-construction}
We derive our corpus from the public 1{,}000-scenario AgentLeak benchmark~\cite{AgentLeak}, which provides tasks, protected values, agent roles, and release policies. We construct long-horizon executions at $T{=}20$; 22 scenarios yield no valid neighboring pair and are excluded, leaving 978. We then retain only \emph{true multi-agent} scenarios, at least two agents, two distinct roles, and observable inter-agent coordination through $C_2$ edges, yielding 485: 138 corporate, 100 finance, 100 healthcare, 147 legal. Each scenario's vault contains multiple protected fields alongside a field-level release policy specifying \texttt{allowed\_set} and \texttt{forbidden\_fields}. We designate a single protected attribute $P$ per scenario as the tracked target for paired-trajectory construction, drawn from \texttt{forbidden\_fields}, with unspecified fields treated as forbidden by default.

For each scenario, we construct a neighboring pair $(\tau,\tau')$ differing only in $P$, evaluated at horizons $T\in\{1,3,5,10,20\}$, fixed within each scenario and held constant across topology and adversary-level conditions to preserve the corpus's horizon distribution. Each of the 485 scenarios is evaluated under six topology cells and three adversary levels, yielding $485 \times 6 \times 3 = 8{,}730$ execution records, each containing the pair and seven observable channels. Detection runs independently per channel and pair member, yielding $7\times2=14$ instances per record and $8{,}730\times14=122{,}220$ total (\autoref{tab:strata-eval}), the unit of channel-level evaluation from which record- and topology-level statistics are computed.

\paragraph{Trace normalization.} We execute each scenario through a custom-built framework-agnostic adapter normalizing traces from all eight orchestration frameworks into a common JSON Lines representation, exposing events on $C_1$--$C_7$ and recording execution and information-flow graphs for attribution and integrity checks. All executions use Qwen3-32B, served locally with vLLM~\cite{vllm}, at temperature 0 with fixed seeds. Tool-call inputs and outputs ($C_3$/$C_4$) come from a deterministic tool layer rather than live services, so paired executions and topology conditions differ only in protected-value substitution and communication structure.

%%%%%%%%%%%%%%%%%%%%%%%%%%%%%%%%%%%%%%%%%%%

\subsection{Baseline Detection Pipeline}\label{subsec:detector}
We adopt AgentLeak's three-tier detection methodology~\cite{AgentLeak}, adapted to our multi-agent traces, applied independently per channel to ask whether $P$ is disclosed, verbatim or via a semantically equivalent expression. Authorization is evaluated separately via the recipient-conditioned policy (\S\ref{sec:threat-model}.

\begin{enumerate}[left=0pt]
\item \textbf{Canary and exact matching (Tier 1).} Deterministically detects verbatim occurrences, checking observable-content capture.
\item \textbf{Structured-field matching (Tier 2).} Deterministic recognizers identify SSNs, Luhn-valid cards, and ICD-10 codes; checksum-validated matches are high-confidence, pattern-only matches are flagged but not counted, and matches are restricted to the target field and its release policy.
\item \textbf{LLM-as-judge (Tier 3).} Resolves remaining instances by evaluating semantic equivalence between channel content and the protected value, using Mistral-Small-24B as judge and Qwen3-32B as generator; a score in $[0,1]$ classifies disclosure when at least the calibrated threshold $\zeta$
(\S\ref{subsec:threshold-calibration}).
\end{enumerate}
\noindent Each disclosure is attributed to the first detecting tier, in order: Tier 1, Tier 2, and Tier 3, preventing duplicate counting and separating exact, structured, and semantic detections.

\subsection{Detection Stratification}\label{subsec:pipeline}
We apply the pipeline to all 122{,}220 instances. A deterministic pre-filter partitions them into three strata (\autoref{tab:strata-eval}): \emph{certain-positive}, resolved by Tier 1 or 2; \emph{certain-negative}, with no protected value introduced into the channel's execution context; and \emph{semantic-middle}, requiring Tier 3 assessment. Tier 3 performance is reported only on the semantic-middle stratum, since combining it with the deterministic strata would conflate exact matching with semantic-judge performance.

\begin{table}[t]
\centering
\caption{\small Stratification of the 122{,}220 evaluation instances.}
\vspace{-0.4cm}
\label{tab:strata-eval}
\small
\tabcolsep = 0.08cm
\begin{tabular}{@{}llrr@{}}
\toprule
Stratum & Resolving mechanism & Instances & Share \\
\midrule
Certain-positive & Tier 1/2 exact or structured match & 70{,}280 & 57.5\% \\
Semantic-middle  & Tier 3 judge & 36{,}568 & 29.9\% \\
Certain-negative & No deterministic signal & 15{,}372 & 12.6\% \\
\midrule
Total & & 122{,}220 & 100.0\% \\
\bottomrule
\end{tabular}
\end{table}

\subsection{Calibration Corpus and Ground Truth}
\label{subsec:calibration-corpus}

We calibrate the Tier-3 threshold on a dedicated corpus of 720 traces generated under the CW-2 topology and $A_0$ attack level, containing 10{,}080 detection instances across two neighboring-pair members and seven channels. A deterministic pass assigns each instance to the three strata defined in~\autoref{tab:strata-eval}; only the semantic-middle stratum is submitted for human annotation. Two domain-expert annotators independently label a stratified 300-instance sample, each instance receiving \textsc{Leak}, \textsc{No-Leak}, or \textsc{Unsure}. Disagreements are adjudicated against a written guideline evaluating disclosure of the designated attribute rather than unrelated sensitive content.  After adjudication, the sample yields 32 leaks and 267 non-leaks among 299 resolved instances. Inter-annotator agreement on raw pre-adjudication labels is substantial, with Cohen's $\kappa = 0.835$
three-way and $0.846$ binary at $95.7\%$ raw agreement, where the three-way figure includes \textsc{Unsure} and the binary figure excludes any instance either annotator so marked, confirming the calibration labels are reliable.

\subsection{Threshold Calibration}
\label{subsec:threshold-calibration}
We calibrate $\zeta$ on the adjudicated semantic-middle sample. The Tier-3 judge operates at temperature 0, producing a deterministic score. We sweep $\zeta\in[0.35,0.95]$ in steps of 0.01, classifying an instance as disclosed when its score is $\geq\zeta$, and selecting the threshold that maximizes $F1$ subject to $\mathrm{FPR}<0.05$. The FPR constraint holds throughout, with a maximum of 0.0112, so selection is determined entirely by $F1$, which is maximized at 0.9231 across $\zeta\in[0.35,0.60)$, an interval containing no calibration score, leaving classification unchanged throughout. We report $\zeta=0.50$ as the representative operating point: precision 0.9091, recall 0.9375, FPR 0.0112, $F1$ 0.9231, values characterizing the semantic-middle population rather than aggregate corpus performance. AgentLeak's~\cite{AgentLeak} threshold, $\zeta=0.72$, yields $F1=0.8525$ on our sample versus 0.9231 at $\zeta=0.50$, so we recalibrate rather than transfer it. As a robustness check, Gemma-3-27B, a different model family, independently scores the labeled sample; at $\zeta=0.50$ the two judges agree on 96.7\% of instances, Cohen's $\kappa=0.846$. Gemma-3-27B is used only for this cross-check and does not affect threshold selection or reported counts.

\begin{table}[t]
\centering
\caption{\small Tier-3 threshold calibration on the main semantic-middle sample
(299 instances, 32 leaks).}
\vspace{-0.4cm}
\label{tab:sweep}
\small
\begin{tabular}{@{}lllllll@{}}
\toprule
$\zeta$ & FPR & FNR & Accuracy & Precision & Recall & $F1$ \\
\midrule
0.35--0.60 & 0.0112 & 0.0625 & 0.9833 & 0.9091 & 0.9375 & 0.9231 \\
0.61--0.70 & 0.0112 & 0.0937 & 0.9799 & 0.9062 & 0.9062 & 0.9062 \\
0.71--0.80 & 0.0112 & 0.1875 & 0.9699 & 0.8965 & 0.8125 & 0.8525 \\
0.81--0.85 & 0.0112 & 0.2187 & 0.9665 & 0.8926 & 0.7812 & 0.8333 \\
0.86--0.90 & 0.0075 & 0.2500 & 0.9665 & 0.9231 & 0.7500 & 0.8276 \\
0.91--0.95 & 0.0037 & 0.3750 & 0.9565 & 0.9524 & 0.6250 & 0.7547 \\
\bottomrule
\end{tabular}
\end{table}

\section{Evaluation}

\subsection{Topology-Dependent Leakage}\label{sec:rq1}
\subsubsection{Experimental Design.}
We evaluate the frozen held-out corpus under all six topology cells and three adversary levels, $A_0$, $A_1$, $A_2$, with detector, judge, prompts, policy, and pipeline fixed throughout (\S\ref{sec:methodology}). Our primary outcome is topology-conditioned new-agent disclosure, $L(\Breve{T}) = P(\text{unauthorized
disclosure} \mid \Breve{T})$, where $\Breve{T}$ denotes topology. We report three metrics:
\begin{enumerate}[leftmargin=*]
\item \textbf{Surface leakage}: the proportion of scenarios with at
least one unauthorized disclosure in any channel.
\item \textbf{New-agent disclosure}: the proportion reaching a
topology-introduced recipient ($\mathit{pos}={pos}_j$), our primary
measure of topology-dependent exposure.
\item \textbf{Unauthorized propagation}: the proportion reaching any
recipient outside the authorized set regardless of position.
\end{enumerate}
\noindent This distinction lets us determine whether topology changes
overall disclosure incidence or primarily where disclosed
information propagates.

\subsubsection{Topology-Conditioned Exposure.}\label{sec:topology-conditioned-exposure}
\begin{figure}
    \centering
    \includegraphics[width=0.8\linewidth]{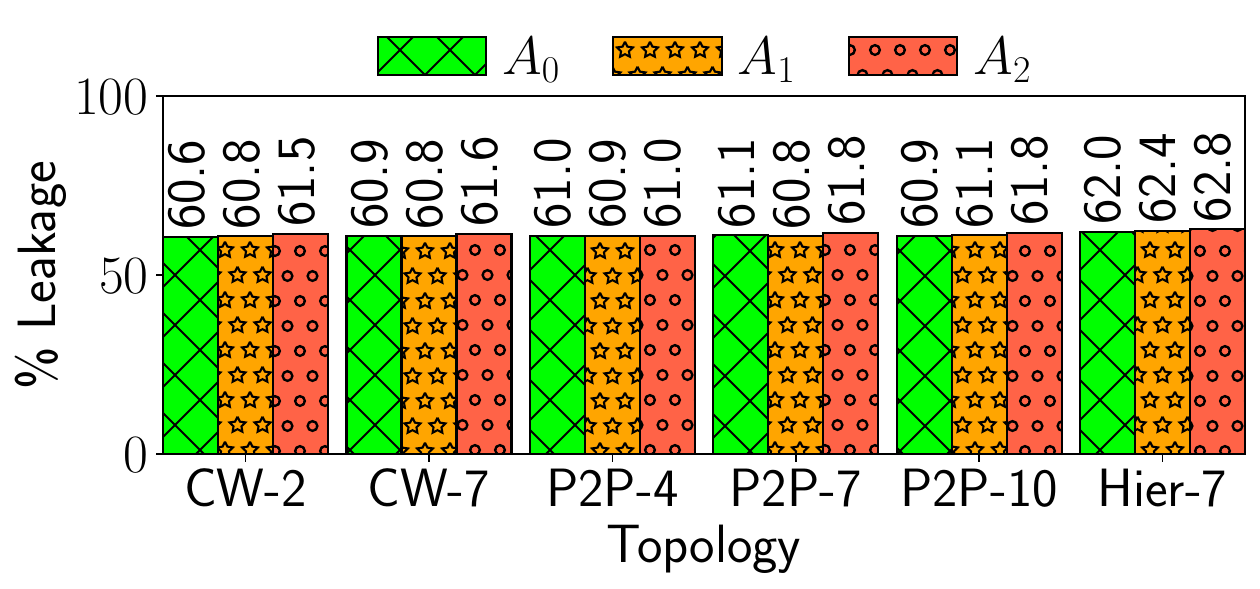}
    \vspace{-0.4cm}
    \caption{\small Surface leakage rate (\%) by topology and adversary level. Each cell contains the proportion of the 485 scenarios exhibiting at least one unauthorized disclosure.}
    \label{fig:rq1-surface}
\end{figure}

\begin{figure}
    \centering
    \includegraphics[width=1.0\linewidth]{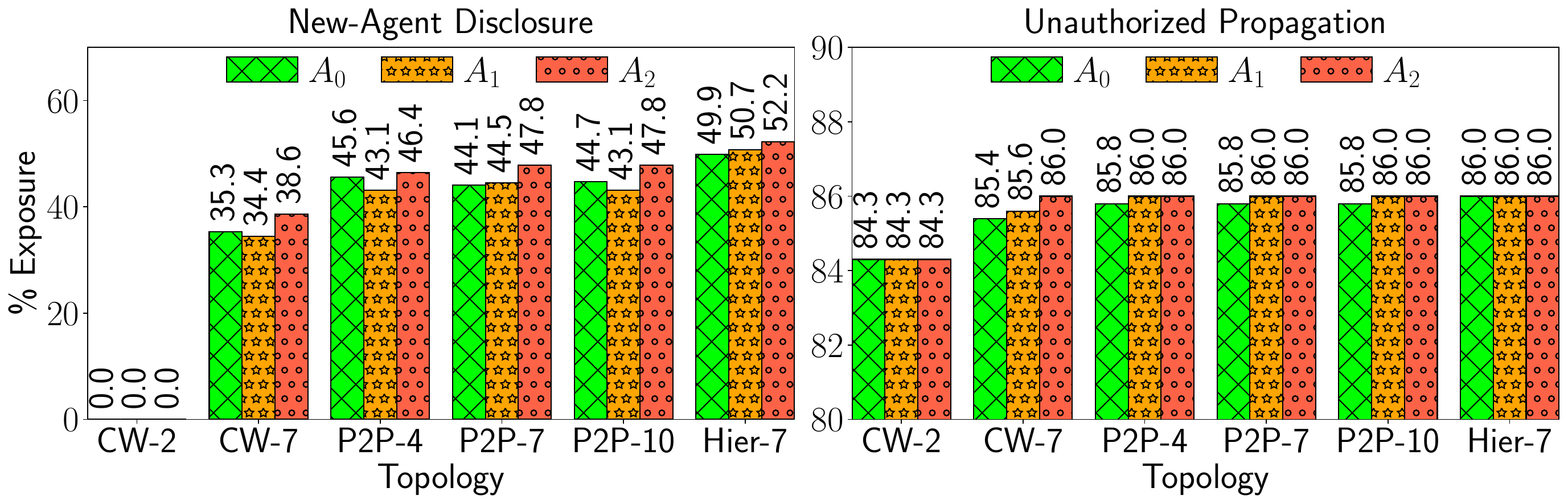}
    \vspace{-0.4cm}
    \caption{\small Recipient-conditioned exposure (\%) by topology and adversary level. New-agent disclosure is the primary topology-conditioned leakage measure $L(T)$. Unauthorized propagation includes any recipient outside the authorized set.}
    \label{fig:rq1-propagation}
\end{figure}

Surface leakage changes little across topologies, remaining between $60.6\%$ and $62.8\%$ across adversary levels (\autoref{fig:rq1-surface}), while recipient-conditioned exposure increases substantially: new-agent disclosure is $0\%$ at CW-2 by construction, $34.4$--$38.6\%$ at CW-7, $43.1$--$47.8\%$ across the P2P cells, and $49.9$--$52.2\%$ at
Hier-7 (\autoref{fig:rq1-propagation}). Unauthorized propagation stays high throughout, from $84.3\%$ at CW-2 to at most $86.0\%$ in richer topologies.
 
% \begin{table}[t]
% \centering
% \small
% \caption{\small Recipient-conditioned exposure (\%) by topology and adversary level. New-agent disclosure is the primary topology-conditioned leakage measure $L(T)$. Unauthorized propagation includes any recipient outside the authorized set.}
% \vspace{-0.4cm}
% \label{tab:rq1-propagation}
% \begin{tabular}{@{}lccc|ccc@{}}
% \toprule
% & \multicolumn{3}{c|}{New-Agent Disclosure}
% & \multicolumn{3}{c}{Unauthorized Propagation} \\\cline{2-7}
% Topology & $A_0$ & $A_1$ & $A_2$
% & $A_0$ & $A_1$ & $A_2$ \\
% \midrule
% CW-2    & 0.0 & 0.0 & 0.0 & 84.3 & 84.3 & 84.3 \\
% CW-7    & 35.3 & 34.4 & 38.6 & 85.4 & 85.6 & 86.0 \\
% P2P-4   & 45.6 & 43.1 & 46.4 & 85.8 & 86.0 & 86.0 \\
% P2P-7   & 44.1 & 44.5 & 47.8 & 85.8 & 86.0 & 86.0 \\
% P2P-10  & 44.7 & 43.1 & 47.8 & 85.8 & 86.0 & 86.0 \\
% Hier-7  & 49.9 & 50.7 & 52.2 & 86.0 & 86.0 & 86.0 \\
% \bottomrule
% \end{tabular}
% \end{table}

At $A_0$, the Wilson $95\%$ confidence intervals for new-agent disclosure are $35.3\%$ [$31.1,39.6$] for CW-7, $45.6\%$ [$41.2,50.0$] for P2P-4, $44.1\%$ [$39.8,48.6$] for P2P-7, $44.7\%$ [$40.4,49.2$] for P2P-10, and $49.9\%$ [$45.5,54.3$] for Hier-7. The corresponding interval for CW-2 is $[0.0,0.8]\%$. We use Cochran's $Q$ test for formal inference because each scenario is evaluated repeatedly across topology conditions (Table~\ref{tab:rq1-stats}). CW-2 is excluded from the topology test for new-agent disclosure because its outcome is structurally constant at zero.
\begin{table}[t]
\centering
\caption{\small Cochran's $Q$ tests for topology and adversary-intensity effects on recipient-based disclosure.}
\vspace{-0.4cm}
\label{tab:rq1-stats}
\small
\begin{tabular}{@{}llrrl@{}}
\toprule
Effect & Metric & $\chi^2$ & $\mathrm{df}$ & $p$ \\
\midrule
Topology & New-agent & 191.8 & 4 & $2\times10^{-40}$ \\
Topology & Propagation & 94.2 & 5 & $9\times10^{-19}$ \\
Adversary & New-agent & 16.0 & 2 & $3\times10^{-4}$ \\
Adversary & Propagation & 9.3 & 2 & $9\times10^{-3}$ \\
\bottomrule
\end{tabular}
\end{table}
All four effects are significant at conventional thresholds (Table~\ref{tab:rq1-stats}). Topology has a larger effect on new-agent disclosure ($\chi^2=191.8$, $\mathrm{df}=4$, $p=2\times10^{-40}$) than on unauthorized propagation ($\chi^2=94.2$, $\mathrm{df}=5$, $p=9\times10^{-19}$): topology moves
new-agent disclosure over a wide range, $0\%$ to $52.2\%$, but propagation only within a narrow 1.7-point band, $84.3\%$ to $86.0\%$, so a smaller absolute range still registers as a strong effect once repeated-measures structure is accounted for. Adversary intensity is also significant but with much smaller statistics, $16.0$ and $9.3$ against $191.8$ and $94.2$, making topology the dominant driver here and adversary intensity secondary.

Hence, the effect concentrates in recipient-conditioned exposure: surface leakage varies by only $2.2$ points across cells, while new-agent disclosure rises from a structural $0\%$ at CW-2 to $52.2\%$ at Hier-7 under $A_2$. Richer topologies form no strict monotonic ordering; P2P-4 exceeds P2P-7 at $A_0$, so results reflect not agent count but communication structure changing which recipients become exposed to already-disclosed information. These disclosures occur through $C_2$ rather than the deterministically injected structured channels, consistent with topology-dependent propagation rather than an artifact of structured-value survival. Overall, this result shows that topology primarily changes the reachable recipient set: richer coordination's principal security consequence is increased unauthorized reach, even as surface disclosure incidence stays nearly unchanged.

\subsection{Compositional Leakage}\label{sec:compositional-results}
We investigate whether disclosure accumulates as an adversary observes progressively longer trajectories, over nested horizons $T_1 \subset T_3 \subset T_5 \subset T_{10} \subset T_{20}$, distinguishing \emph{cumulative disclosure}, distinct protected values disclosed by horizon $T_k$, from \emph{marginal disclosure}, those first disclosed between consecutive horizons, and the primary measure of compositional leakage: sustained disclosure at later horizons would indicate accumulation, concentration at early horizons front-loading. We evaluate behavioral channels ($C_1$/$C_2$/$C_6$/$C_7$), all channels ($C_1$--$C_7$), and tool channels ($C_3$/$C_5$), including only agent-authored events. Each turn beyond the opening task is scripted as one of eight \emph{interaction archetypes} (\autoref{tab:archetype-defs}), reflecting a distinct conversational function.

\paragraph{Disclosure is predominantly front-loaded.}
Cumulative disclosure increases with horizon, but most first-time disclosures occur at the first transition (Table~\ref{tab:comp-cumulative}). For behavioral channels, mean cumulative disclosure rises from $0.020$ at $T_1$ to $0.321$ at $T_{20}$; of $311$ first-time disclosures across $970$ trajectories, $188$ ($60.5\%$) occur at $T_1{\to}T_3$, versus $16$ at $T_3{\to}T_5$, $19$ at $T_5{\to}T_{10}$, and $69$ at $T_{10}{\to}T_{20}$ (Table~\ref{tab:comp-marginal}). The same pattern holds across all channels and tool channels: $62.2\%$
and $64.0\%$ of first-time disclosures, respectively, occurring at $T_1{\to}T_3$: longer horizons increase cumulative exposure, but later turns add little new disclosure, and we find no evidence of sustained accumulation.

\begin{table}[t]
\centering
\caption{\small Cumulative disclosure: mean number of distinct protected values disclosed per trajectory across nested horizons. Values are judge-scored and pooled over $\tau/\tau'$ ($n=970$).}
\vspace{-0.4cm}
\label{tab:comp-cumulative}
\small
\begin{tabular}{@{}lccccc@{}}
\toprule
Scope & $T_1$ & $T_3$ & $T_5$ & $T_{10}$ & $T_{20}$ \\
\midrule
Behavioral ($C_1$/$C_2$/$C_6$/$C_7$) & 0.020 & 0.213 & 0.230 & 0.250 & 0.321 \\
All channels ($C_1$--$C_7$)          & 0.020 & 0.237 & 0.251 & 0.281 & 0.349 \\
Tool only ($C_3$/$C_5$)              & 0.000 & 0.057 & 0.057 & 0.081 & 0.089 \\
\bottomrule
\end{tabular}
\end{table}

\begin{table}[t]
\centering
\caption{\small Marginal first-time disclosures by horizon transition. Counts denote trajectories with a newly disclosed protected value among $970$ trajectories; re-disclosures are excluded.}
\vspace{-0.4cm}
\label{tab:comp-marginal}
\small
\begin{tabular}{@{}lrrrrr@{}}
\toprule
Scope & $T_1$ & $T_1{\to}T_3$ & $T_3{\to}T_5$ &
$T_5{\to}T_{10}$ & $T_{10}{\to}T_{20}$ \\
\midrule
Behavioral ($C_1$/$C_2$/$C_6$/$C_7$) & 19 & \textbf{188} & 16 & 19 & 69 \\
All channels ($C_1$--$C_7$)          & 19 & \textbf{211} & 13 & 30 & 66 \\
Tool only ($C_3$/$C_5$)              & 0  & \textbf{55}  & 0  & 24 & 7  \\
\bottomrule
\end{tabular}
\end{table}

\paragraph{Confirmation probes dominate disclosure.}
Confirmation probes (archetype G) account for $574$ of $2{,}212$ disclosing turns across all channels, a $25.9\%$ rate (Table~\ref{tab:comp-archetype}), versus $0.9\%$ to $2.7\%$ for other archetypes, the dominant disclosure context though not exclusive: $14.7\%$ of first-time disclosures occur outside it, in handoffs, elaboration, justification, and closeout turns.

\begin{table}[t]
\centering
\caption{\small Disclosure by interaction archetype in the all-channel scope.
Rate is the fraction of turns of each archetype that disclose the
protected value.}
\vspace{-0.4cm}
\label{tab:comp-archetype}
\small
\begin{tabular}{@{}lrrr@{}}
\toprule
Archetype & Disclosures & Turns & Rate \\
\midrule
G (confirmation probe) & 574 & 2{,}212 & \textbf{0.259} \\
F (rapport)             & 78  & 2{,}916 & 0.027 \\
B (handoff)              & 97  & 4{,}101 & 0.024 \\
E (closeout)             & 62  & 2{,}643 & 0.023 \\
C (justification)        & 67  & 3{,}761 & 0.018 \\
BASE (initial task)      & 19  & 1{,}777 & 0.011 \\
D (error handling)       & 36  & 3{,}699 & 0.010 \\
A (elaboration)          & 63  & 6{,}646 & 0.009 \\
\bottomrule
\end{tabular}
\end{table}

\paragraph{Tool channels provide an independent leakage surface.}
Tool channels produce fewer disclosures than behavioral ones but expose a distinct surface (\autoref{tab:comp-cumulative} and \autoref{tab:comp-marginal}): mean $T_{20}$ disclosure through $C_3$/$C_5$ is $0.089$ versus $0.321$ behaviorally, and is similarly front-loaded, $55$ of $86$ first-time tool disclosures occurring at $T_1{\to}T_3$, meaning communication-only auditing would miss disclosures in tool-call arguments and memory writes.

\subsection{Reconstruction: No Confirmed Emergent Recovery} \label{subsec:reconstruction-results}
Direct disclosure propagates once emitted (\S\ref{sec:rq1}--\ref{sec:compositional-results}); we now test the more serious threat of \emph{reconstruction}, an adversary recovering $P$ from evidence it was never explicitly given.

\subparagraph{\textbf{Inference.}} Whether $P$ is recoverable from contextual cues within a single non-disclosing channel, across 2{,}954 eligible instances in four domains, yields no confirmed reconstruction, $0/2{,}954$ (\autoref{tab:recon-summary}). The judge produced 22 raw positives, all rejected on audit, attributable to decoy-value selection (14), field-type confusion (6), and generic
topical over-attribution (2).

\subparagraph{\textbf{Cross-channel.}} Whether two independently non-disclosing channels jointly reconstruct $P$, across 3{,}195 eligible pairs in all four domains, confirms no cross-channel linkage, $0/3{,}195$ (\autoref{tab:recon-summary}). 25 raw positives (0.78\%) were rejected on audit; typically, two channels independently contained the same incorrect candidate value rather than jointly recovering the true
one.

\subparagraph{\textbf{Cross-agent.}} Whether pooling two agents' views reconstructs $P$ when neither individually possesses it, over all 172{,}949 eligible pairs across six topology cells, yields $0$ confirmed reconstructions (\autoref{tab:recon-summary}). 503 raw positives (0.29\%) were rejected: 443 showed no evidence of the true value, and 60 contained it verbatim in a single agent's view, possession rather than linkage.

\begin{table}[t]
\centering
\small
\caption{\small Reconstruction results by aggregation axis. All positives were
manually audited; emergent recovery is zero on every axis.}
\vspace{-0.4cm}
\label{tab:recon-summary}
\begin{tabular}{@{}lllll@{}}
\toprule
Axis & Eligible & Raw positives & \texttt{Single\_src} & Audited \\
\midrule
Inference     & 2{,}954   & 22  &  39.5\% & \textbf{0} \\
Cross-channel & 3{,}195   & 25  &  33.0\% & \textbf{0} \\
Cross-agent   & 172{,}949 & 503 & 87.2\% & \textbf{0} \\
\bottomrule
\end{tabular}
\end{table}

Since emergent recovery is zero on every axis, cumulative recoverability reduces to the single-source rate; we do not report this separately. A complementary turn-level view (\S\ref{sec:compositional-results}) shows first-time disclosures are
front-loaded rather than accumulating, 211 of 339 ($62\%$) occurring at the first transition, $T_1{\to}T_3$, describing timing of direct disclosure, not reconstruction. Across 179{,}098 exhaustively evaluated instances and pairs, every raw positive manually audited, we find no confirmed evidence a withheld attribute is reconstructed from non-disclosing content, alone or combined, under the evaluated corpus and protocol, though this does not establish absence under
other corpora or strategies.

\subsection{Behavioral Leakage} \label{sec:behavioral-results}
A distinct threat is $P$ influencing observable behavior without being disclosed. We measure this over 33 structural features per trajectory encoding nothing of $P$: pairwise distinguishability between neighboring executions, and population-level prediction of $P$'s semantic category.

\subsubsection{Pair Distinguishability} \label{subsec:pair-dist-results}

\subparagraph{\textbf{Method.}} We identify 244 matched pairs $(\tau,\tau')$ differing only in $P$, with no judge-confirmed disclosure on either side (488 trajectories); a logistic-regression distinguisher predicts $\tau$ versus $\tau'$ from behavioral features under scenario-grouped cross-validation, with distinguishing advantage as $\text{accuracy}-0.5$ against a shuffled-label null per horizon.

\noindent\textbf{Results.} The advantage is transient, not monotonic, near-random at $T_1$ ($50.2\%$), peaking at $T_5$ ($56.6\%$), near-random again at $T_{20}$ ($50.0\%$). The $T_5$ result is statistically distinguishable from the null ($z\approx12.8$), though distinguishability alone does not establish a $P$-dependent side-channel, tested directly below.

\begin{table}[t]
\centering
\small
\caption{\small Pairwise distinguishability by horizon. The $T_5$ signal does not
survive the confound analysis below.}
\vspace{-0.4cm}
\label{tab:pairdist}
\begin{tabular}{@{}lrrr@{}}
\toprule
Horizon & Real accuracy & Real advantage & Shuffled advantage \\
\midrule
$T_1$    & 0.502 & $+0.002$ & $-0.000$ \\
$T_3$    & 0.531 & $+0.031$ & $-0.008$ \\
$T_5$    & 0.566 & $+0.066$ & $-0.002$ \\
$T_{10}$ & 0.537 & $+0.037$ & $-0.000$ \\
$T_{20}$ & 0.500 & $+0.000$ & $-0.004$ \\
\bottomrule
\end{tabular}
\end{table}

\noindent\textbf{Confound analysis.} Confound analysis attributes the $T_5$ signal to corpus construction, not $P$, on three grounds. A genuine fingerprint should not favor one pair member arbitrarily assigned; four of the eleven strongest $T_5$ features show directional bias ($|t|\geq2$, e.g.\ \texttt{tool\_calls\_archetype\_A} favoring $\tau$ in 8/244 pairs and $\tau'$ in none). Generation order is perfectly confounded with pair label, $\tau$ always generated before $\tau'$ ($\mathrm{corr}=1.000$), leaving no within-pair variation to isolate an order effect. And the protected value itself does not explain the difference: value-length and response-length correlate at only $r=0.06$, yet equal-length values (e.g.\ \texttt{63000} vs.\ \texttt{75267}) still produce $\sim\!100$-character response differences.

As an order-invariant alternative, comparing true pairs against domain-constrained synthetic re-pairings yields larger advantages (0.21--0.37, Table~\ref{tab:pairdist-confound}) than the original 0.066, with all 18 non-trivial features showing smaller within-pair differences for real than synthetic pairs (ratios 0.13--0.69), consistent with scenario identity and task similarity rather than a targeted fingerprint. A definitive test would require randomized generation order or repeated generations per scenario; without that, we treat $T_5$ as confounded, not behavioral leakage.

\begin{table}[t]
\centering
\small
\caption{\small Order-invariant control: real matched pairs vs.\ synthetic
re-pairings.}
\vspace{-0.4cm}
\label{tab:pairdist-confound}
\begin{tabular}{@{}lrr@{}}
\toprule
Horizon & Real advantage & Shuffled null \\
\midrule
$T_1$    & $+0.367$ & $-0.018$ \\
$T_3$    & $+0.346$ & $-0.008$ \\
$T_5$    & $+0.342$ & $-0.015$ \\
$T_{10}$ & $+0.338$ & $-0.007$ \\
$T_{20}$ & $+0.207$ & $-0.011$ \\
\bottomrule
\end{tabular}
\end{table}

\subsection{Synthesis: Leakage Is Primarily Direct and Propagative.}
\label{sec:leakage-synthesis}
Explicit disclosure propagates across agents, channels, and turns; all three reconstruction axes yield zero confirmed emergent recovery across 179{,}098 exhaustively evaluated instances and pairs, every raw positive manually audited, and behavioral analysis finds no persistent value-level fingerprint. The dominant leakage pathway is thus direct disclosure followed by propagation: richer topology expands the recipient set for an already-disclosed value, and longer executions add further disclosure opportunities. These findings motivate defenses constraining initial disclosure and subsequent propagation (\S\ref{sec:sanitizer}).

\subsection{Cross-Framework Leakage Measurement}\label{sec:cross-framework}

This experiment evaluates whether orchestration framework choice affects privacy leakage under an identical corpus, generator, authorization policy, and detection pipeline.

\subsubsection{Experimental Design.}\label{subsec:cf-design}
We evaluate the same 485 scenarios across eight frameworks: AutoGen~\cite{AutoGen}, CrewAI~\cite{Crewai}, Google ADK~\cite{Google_ADK}, LangChain~\cite{Langchain}, LangGraph~\cite{Langgraph}, LlamaIndex~\cite{LlamaIndex_github}, OpenAI Agents SDK~\cite{Openai_agents_sdk}, and Semantic Kernel~\cite{Semantic_kernel}, each scenario runs as an attribute-neighboring pair $(\tau,\tau')$, yielding $485 \times 8 \times 2 = 7{,}760$ executions. All use the same generator (Qwen3-32B via vLLM), deterministic decoding, identical prompts, tools, and attribute assignments, with events collected over $C_1$--$C_7$ and normalized into the common trace schema introduced in \S\ref{subsec:corpus-construction}, totaling $214{,}853$ assessable channel events across frameworks. Each framework runs through its native coordination abstraction, delegation, handoffs, workflows, or group chat, rather than a framework-independent layer, preserving the communication behavior that is the object of comparison.

\subsubsection{Detection and Authorization.}\label{subsec:cf-detection}
Leakage detection uses the frozen three-tier pipeline (\S\ref{subsec:detector}); the judge is identical across frameworks and never evaluates its own model family's outputs. Detection and authorization are evaluated separately: the judge decides only whether the attribute appears, and authorization is applied afterward against the field-level release policy $\mathcal{A}(\breve{a}, \varsigma, c, r, \mathit{pos})$ (\S\ref{sec:threat-model}). We report: (1) \textit{Raw leakage}: all semantic detections; (2) \textit{Policy-filtered leakage}: detections excluding disclosures explicitly authorized by the scenario policy. The latter corresponds to the contextual-integrity violation used for all comparisons below.

\subsubsection{Value Fidelity Verification.}\label{subsec:cf-fidelity}
Since detection does not verify the disclosed value is correct, we measure value fidelity independently for canonical attributes (identifiers, phone numbers, dates of birth, account numbers, canary tokens, names), comparing generated values against ground-truth values after canonicalization, classifying each as \textit{exact}, \textit{corrupted}, or \textit{substituted}, reported independently to rule out framework rankings explained by confabulation.

\subsubsection{Framework Leakage Rates.}\label{subsec:cf-results}
Table~\ref{tab:framework-leakage} reports raw leakage, policy-filtered leakage, and value fidelity per framework, sorted by policy-filtered leakage.

\begin{table}[t]
\centering
\caption{\small Leakage and value fidelity across eight orchestration frameworks,
sorted by policy-filtered leakage.}
\vspace{-0.4cm}
\small
\tabcolsep = 0.08cm
\begin{tabular}{lccc}
\toprule
Framework & Raw leakage & Policy-filtered leakage & Value fidelity \\
\midrule
AutoGen           & 43.85\% & 43.38\% & 87.15\% \\
LangChain         & 25.54\% & 25.50\% & 25.59\% \\
LangGraph         & 23.75\% & 23.71\% & 26.40\% \\
CrewAI            & 21.61\% & 21.01\% & 27.75\% \\
Semantic Kernel   & 1.56\%  & 0.67\%  & 67.11\% \\
LlamaIndex        & 0.72\%  & 0.50\%  & 36.23\% \\
OpenAI Agents SDK & 1.56\%  & 0.48\%  & 42.55\% \\
Google ADK        & 0.80\%  & 0.29\%  & 47.06\% \\
\midrule
Overall           & 21.63\% & 21.44\% & 48.47\% \\
\bottomrule
\end{tabular}
\label{tab:framework-leakage}
\end{table}

The frameworks separate into two groups: AutoGen, LangChain, LangGraph, and CrewAI between $21.0\%$ and $43.4\%$ filtered leakage; Semantic Kernel, LlamaIndex, OpenAI Agents SDK, and Google ADK below $0.7\%$. The gap between groups, CrewAI $21.01\%$ to Semantic Kernel $0.67\%$, is roughly $31\times$, and the overall extremes, AutoGen $43.38\%$ versus Google ADK $0.29\%$, roughly
$150\times$. This two-group separation is the reliable finding, not the ordering \emph{within} each group: \S\ref{subsec:cf-robustness} establishes a cross-corpus noise floor no larger than $0.41$ points, and every low-group gap, $0.02$--$0.19$pp, falls well below it, so those four should be read as tied. The high-group's internal gaps, $1.79$--$17.88$pp, exceed the floor by an order of magnitude and are reliably ordered. The authorization filter changes overall leakage by only $0.19$ points, and the two-group separation is unchanged
after filtering.

\begin{figure}
    \centering
    \includegraphics[width=0.75\linewidth]{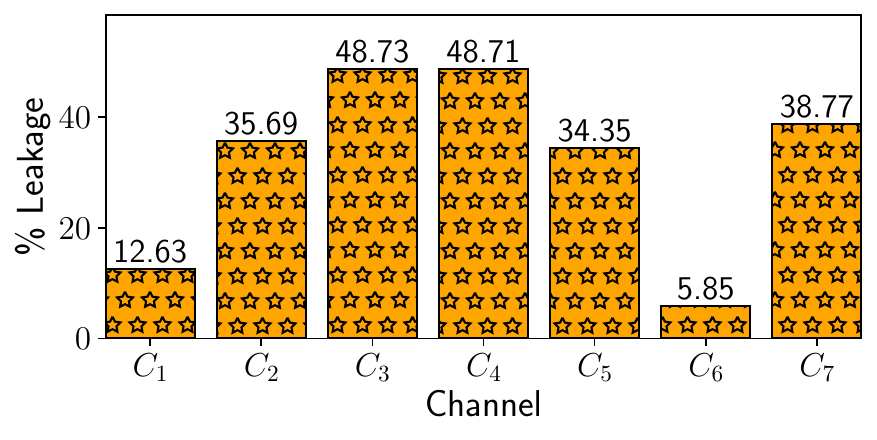}
    \vspace{-0.5cm}
    \caption{\small Leakage rates across observable channels on all frameworks.}
    \label{fig:framework-channel-leakages}
\end{figure}

As shown in \autoref{fig:framework-channel-leakages}, leakage concentrates in non-user-facing channels: tool inputs ($C_3$) and outputs ($C_4$) exhibit the highest rates, followed by artifacts ($C_7$), inter-agent communication ($C_2$), and shared memory ($C_5$), while user-facing outputs ($C_1$) account for substantially
fewer disclosures, showing that evaluating only final responses understates privacy exposure in MALS.

\subsubsection{Value Fidelity Is Independent of Leakage.}\label{subsec:cf-fidelity-results}
Leakage frequency and value fidelity measure different properties. AutoGen leaks most frequently while reproducing values most accurately ($87.15\%$); LlamaIndex leaks rarely but correctly only $36.23\%$ of verifiable cases, so low leakage can reflect avoided disclosure or incorrect and substituted values, fidelity requiring independent measurement.

To rule out confabulation as the source of framework ranking, we joined per-instance fidelity with leakage decisions: exact-value disclosures were almost always detected, substituted values were never classified as valid, and the two-group separation persists when restricted to exact-value disclosures, so the grouping is not explained by confabulation.

\subsubsection{Robustness Across Generation Regimes.}\label{subsec:cf-robustness}
We evaluate robustness using three additional corpora from the identical scenario pool and configurations under temperatures $0.2$, $0.5$, and $0.7$, each unseeded, comparing policy-filtered leakage against the temperature-$0.0$ corpus of record
(Table~\ref{tab:framework-robustness}).

\begin{table}[t]
\centering
\caption{\small Policy-filtered leakage (\%) across four independently generated
corpora. Deltas are relative to the temperature-$0.0$ corpus of record.}
\vspace{-0.4cm}
\tabcolsep = 0.06cm
\label{tab:framework-robustness}
\small
\begin{tabular}{lrrrr}
\toprule
Framework & Temp.\ 0.0 & Temp.\ 0.2 ($\Delta$) & Temp.\ 0.5 ($\Delta$) & Temp.\ 0.7 ($\Delta$) \\
\midrule
AutoGen           & 43.38 & 43.34 ($-$0.04) & 43.41 (+0.03) & 43.25 ($-$0.13) \\
LangChain         & 25.50 & 25.59 (+0.09)   & 25.69 (+0.19) & 25.87 (+0.37)   \\
LangGraph         & 23.71 & 23.79 (+0.08)   & 23.90 (+0.19) & 23.93 (+0.22)   \\
CrewAI            & 21.01 & 21.40 (+0.39)   & 21.04 (+0.03) & 21.41 (+0.40)   \\
Semantic Kernel   & 0.67  & 0.82 (+0.15)    & 0.82 (+0.15)  & 0.84 (+0.17)    \\
LlamaIndex        & 0.50  & 0.45 ($-$0.05)  & 0.49 ($-$0.01) & 0.41 ($-$0.09)  \\
OpenAI Agents SDK & 0.48  & 0.50 (+0.02)    & 0.54 (+0.06)  & 0.52 (+0.04)    \\
Google ADK        & 0.29  & 0.37 (+0.08)    & 0.32 (+0.03)  & 0.40 (+0.11)    \\
\midrule
Overall           & 21.44 & 21.53 (+0.09)   & 21.59 (+0.15) & 21.67 (+0.23)   \\
\bottomrule
\end{tabular}
\end{table}

Framework-level leakage is stable across the range: the high- and low-leakage groups remain separated by more than $19.8$ points at every temperature, with no framework changing group membership. Maximum per-framework drift from the temperature-$0.0$ anchor is small throughout ($0.39$pp at $0.2$, $0.21$pp at $0.5$, $0.41$pp at $0.7$) but non-monotonic, temperature $0.5$ being a local minimum;
the framework responsible for the maximum shifts too, CrewAI at $0.2$ and $0.7$, LangChain at $0.5$, since CrewAI's own drift briefly turns negligible and negative at $0.5$, attributed below to CrewAI specifically rather than temperature.

The apparent low-leakage-group instability noted earlier does not hold under the complete four-regime picture. The raw metric gives an identical ordering: Semantic Kernel, OpenAI Agents SDK, Google ADK, LlamaIndex, at all four temperatures, no raw-metric swap ever occurring. The policy-filtered ordering shows one inversion,
LlamaIndex above OpenAI Agents SDK, only in the fixed-seed temperature-$0.0$ corpus; the three unseeded regimes independently agree on Semantic Kernel, OpenAI Agents SDK, LlamaIndex, Google ADK, so the inversion is best read as specific to that corpus rather than an instability recurring under resampling. All four low-leakage frameworks remain far below the high-leakage group's own internal noise floor regardless of metric or regime, and should be read as tied rather than ranked.

Under fixed temperature and seed, event content reproduces identically in $83.99\%$ of cases at temperature $0.0$, the residual reflecting floating-point non-determinism rather than sampling and not eliminable by seeding; reproducibility does not degrade at higher temperature ($89.92\%$ at $0.5$, $89.68\%$ at $0.7$). CrewAI accounts for nearly all residual non-reproducibility at every temperature ($3/16$, $3/16$, $4/16$ mismatched pairs at $0.0$, $0.5$, $0.7$, against $0/16$ for LangChain and LangGraph throughout), consistent with driving the non-monotonic drift above; framework-level aggregate rates remain stable regardless, indicating the effect reflects aggregate behavior over hundreds of trials rather than any
single generation path. This claim establishes stability under generation variance across temperatures $0.0$ through $0.7$ only; it does not address scenario-sampling, judge-choice, or threshold-recalibration sensitivity, and the non-monotonic temperature relationship should not be extrapolated beyond the tested range.

\subsection{Model Generalization}\label{subsec:model-generalization}
The topology gradient in \S\ref{sec:rq1} was initially measured using \textsc{Qwen3-32B}. To assess generalization across generators, we repeat the sweep with three additional models from distinct families, \textsc{GPT-5-mini}, \textsc{Gemini-3.7-Flash}, and \textsc{Llama-3.3-70B}, holding framework, corpus, topology, pipeline, and detector fixed across all 485 scenarios and six cells. This separates whether topology-induced exposure occurs from how frequently it does. The former is consistent across generators, whereas the latter varies substantially, with cell ordering unpreserved across models.

\subsubsection{Topology-induced exposure generalizes, but its magnitude varies.}
\label{subsubsec:model-gradient}
\begin{figure}
    \centering
    \includegraphics[width=1.0\linewidth]{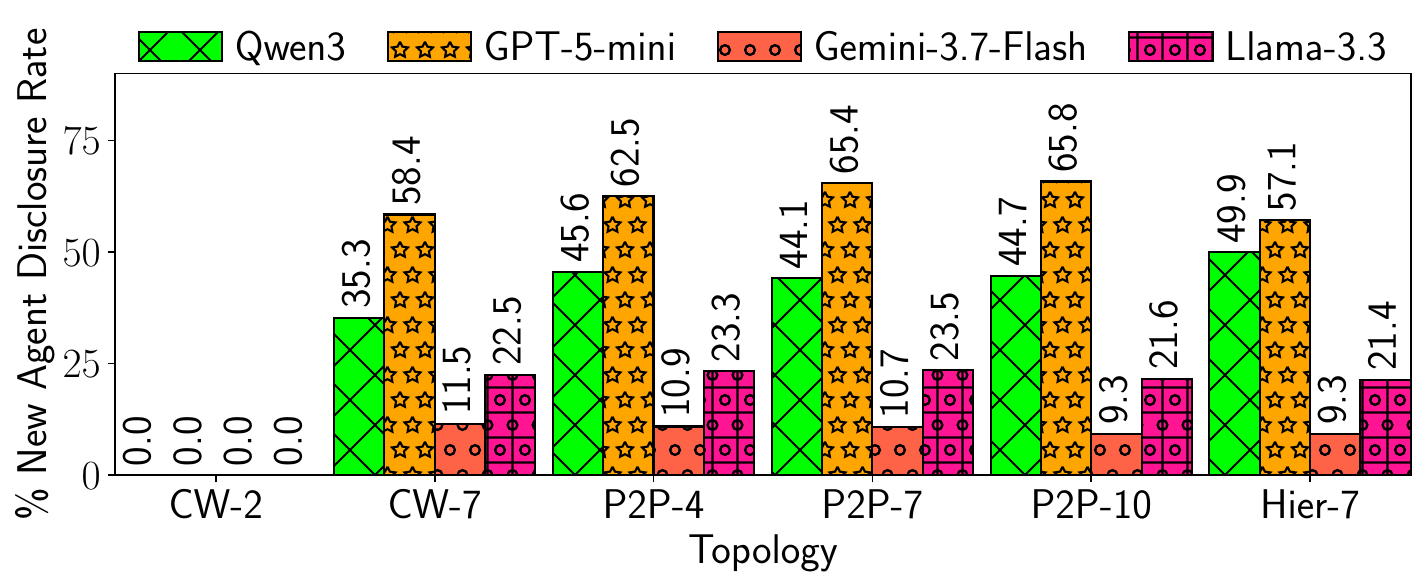}
    \vspace{-0.8cm}
    \caption{\small New-agent disclosure rate by topology cell across four generator models under LangGraph ($n=485$ scenarios per cell). Exposure magnitude and cell ordering vary across generators in richer topology cells.}
    \label{fig:model-gradient}
\end{figure}

Topology-induced exposure occurs for every generator (\autoref{fig:model-gradient}). CW-2 yields no detected disclosure for any model, while every richer topology yields a non-zero rate for all four; the smallest, $9.3\%$ for \textsc{Gemini-3.7-Flash}, remains separated from CW-2's confidence bound, so the effect is not specific to \textsc{Qwen3-32B}. The magnitude differs substantially across generators. Over richer cells, \textsc{GPT-5-mini} ranges $57.1$--$65.8\%$, \textsc{Qwen3-32B} $35.3$--$49.9\%$, \textsc{Llama-3.3-70B} $21.4$--$23.5\%$, and
\textsc{Gemini-3.7-Flash} $9.3$--$11.5\%$, so identical nominal structures yield very different rates by generator (\autoref{fig:model-gradient}).

% \begin{table}[t]
% \centering
% \small
% \caption{\small New-agent disclosure rate by topology cell across four generator models under LangGraph ($n=485$ scenarios per cell). All richer topology
% cells exhibit non-zero exposure for every generator, while exposure
% magnitude and cell ordering vary across generators.}
% \vspace{-0.4cm}
% \label{tab:model-gradient}
% \begin{tabular}{lrrrr}
% \toprule
% Cell & Qwen3 & GPT-5-mini & Gemini & Llama \\
% \midrule
% CW-2   & 0.0\%  & 0.0\%  & 0.0\% & 0.0\% \\
% CW-7   & 35.3\% & 58.4\% & 11.5\% & 22.5\% \\
% P2P-4  & 45.6\% & 62.5\% & 10.9\% & 23.3\% \\
% P2P-7  & 44.1\% & 65.4\% & 10.7\% & 23.5\% \\
% P2P-10 & 44.7\% & 65.8\% & 9.3\%  & 21.6\% \\
% Hier-7 & \textbf{49.9\%} & 57.1\% & 9.3\% & 21.4\% \\
% \bottomrule
% \end{tabular}
% \end{table}

Ordering also varies: \textsc{Qwen3-32B} peaks at Hier-7 ($49.9\%$), while the other three instead rise sharply from CW-2 and then vary narrowly. These results are unsupportive of any model-independent topology-to-rate mapping.  The highest-to-lowest richer-cell gap, $65.8\%$ versus $9.3\%$, roughly sevenfold, is reported descriptively: \textsc{GPT-5-mini} shows generation variability even at temperature zero, so small differences among its cells are not a definitive ranking. The robust
observation is separation between model-level ranges, not internal ordering within any one model.

\subsubsection{Model behavior contributes to exposure variation.} \label{subsubsec:model-disclosure-behavior}

\textsc{Gemini-3.7-Flash}'s comparatively low, flat gradient motivated closer inspection of its Hier-7 executions. A manual recheck of sampled non-disclosing runs found no missed disclosures to topology-introduced agents; judge scores were $0.0$ for $87\%$ of examined events, well below $\zeta=0.50$, and topology construction was verified, giving no evidence of construction failure or a
threshold artifact.

Residual exposure instead tracks attribute type (Table~\ref{tab:gemini-attribute}): structured identifiers (SSN, date of birth, phone, email, account number, salary, balance, credit score) are rarely forwarded ($0.0$--$3.7\%$), while contextual and
narrative attributes reach far higher rates, confidential strategy $27.6\%$, internal notes $36.4\%$, criminal history $63.6\%$, consistent with the flat cross-topology pattern and illustrating why an aggregate rate can conceal which information types actually propagate; we do not interpret this as a universal model property.

% \begin{table}[t]
% \centering
% \small
% \caption{\small \textsc{Gemini-3.7-Flash} Hier-7 disclosure rates by protected
% attribute type. Structured identifiers exhibit lower observed disclosure
% rates than contextual and narrative attributes.}
% \vspace{-0.4cm}
% \label{tab:gemini-attribute}
% \begin{tabular}{lr}
% \toprule
% Attribute & Disclosure rate \\
% \midrule
% SSN / DOB / phone / email / account & 0.0\% \\
% Salary / balance / credit score     & 0.0\% \\
% Address                             & 0.9\% \\
% Settlement amount                   & 3.7\% \\
% \midrule
% Confidential strategy               & 27.6\% \\
% Internal notes                      & 36.4\% \\
% Criminal history                    & 63.6\% \\
% \bottomrule
% \end{tabular}
% \end{table}

\subsubsection{Implications for Topology-Aware Privacy Enforcement.}
\label{subsubsec:model-generalization-implication}
The model sweep adds a second generalization dimension beyond the framework comparison in \S\ref{sec:cross-framework}: nominally equivalent topologies expose differently under different orchestration semantics, and the same topology and framework expose differently across generators. Neither topology nor generator
identity alone establishes that a flow is authorized, since one generator's high exposure under a topology can be another's low exposure, and a generator suppressing structured identifiers may still propagate contextual ones. RECONS addresses this at the point of detected disclosure rather than by assumption, evaluating realized recipient and execution context before preserving, transforming, or suppressing a value (\S\ref{sec:sanitizer}). Since propagation depends on both orchestration and generation, this result reinforces recipient-conditioned enforcement on realized information flow over static assumptions about topology or generator.

\subsection{Framework Generalization}\label{subsec:framework-generalization}
Topology-dependent exposure observed earlier (\S\ref{sec:rq1}) may depend on LangGraph's execution semantics, so we replicate the full sweep in AutoGen, a message-passing framework, holding the generator fixed to \textsc{Qwen3-32B} from the same endpoint and reusing the
framework-agnostic pipeline and frozen detector, across all 485 scenarios in each of six cells, $2{,}910$ scenario-cell executions
total. This addresses whether topology-introduced exposure persists across frameworks, whether its magnitude is comparable, and whether the topology-specific pattern is preserved: the first is affirmed, the latter two are not, with exposure occurring in both frameworks but differing in magnitude and distribution across cells.

\subsubsection{Cross-Framework Replication.}\label{subsubsec:fw-replication}
\begin{figure}[h]
    \centering
    \includegraphics[width=0.75\linewidth]{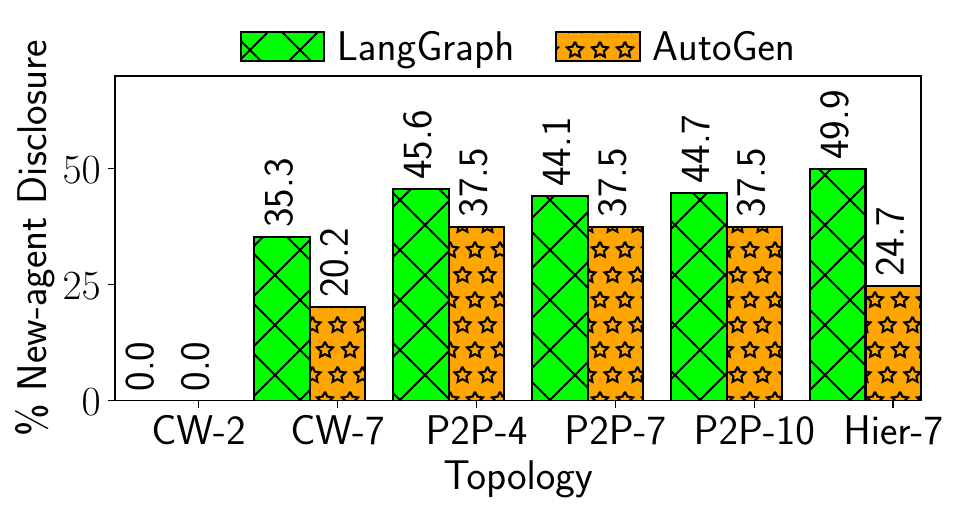}
    \vspace{-0.5cm}
    \caption{\small New-agent disclosure rate by topology cell for LangGraph and AutoGen over the full corpus ($n=485$ per cell), using the same Qwen3-32B generator. Topology-introduced exposure occurs in every richer cell in both frameworks, but its magnitude and topology-specific ordering differ.}
    \label{fig:framework-gradient}
\end{figure}

\subparagraph{Exposure is observed in every richer cell in both
frameworks (\autoref{fig:framework-gradient})}. CW-2 is a structural anchor with no topology-introduced recipient, so its rate is $0\%$ in both, a consequence of the metric definition rather than freedom from other disclosures. For richer cells, AutoGen ranges $20.2$--$37.5\%$ versus LangGraph's $35.3$--$49.9\%$, so the effect is not LangGraph-specific, though magnitude and ordering are not preserved: LangGraph peaks at Hier-7 ($49.9\%$), while AutoGen peaks in the P2P cells ($37.5\%$), with Hier-7 substantially lower ($24.7\%$). Thus, nominal topology does not uniquely determine exposure magnitude; the same topology yields different rates under different orchestration semantics.

% \begin{table}[t]
% \centering
% \small
% \caption{\small New-agent disclosure rate by topology cell for LangGraph and AutoGen over the full corpus ($n=485$ per cell), using the same Qwen3-32B generator. Topology-introduced exposure occurs in every richer cell in both frameworks, but its magnitude and topology-specific ordering differ.}
% \vspace{-0.4cm}
% \label{tab:framework-gradient}
% \begin{tabular}{lrrl}
% \toprule
% Cell & LangGraph & AutoGen & Interpretation \\
% \midrule
% CW-2   & 0.0\%  & 0.0\%  & structural anchor \\
% CW-7   & 35.3\% & 20.2\% & richer topology \\
% P2P-4  & 45.6\% & 37.5\% & richer topology \\
% P2P-7  & 44.1\% & 37.5\% & richer topology \\
% P2P-10 & 44.7\% & 37.5\% & richer topology \\
% Hier-7 & \textbf{49.9\%} & 24.7\% & ordering differs \\
% \bottomrule
% \end{tabular}
% \end{table}

\subsubsection{Realized-Path Analysis.}\label{subsubsec:fw-realized-path}
The cross-framework differences suggest nominal topology does not fully determine realized propagation, so we examine whether the exposure gap tracks the realized communication path, focusing on Hier-7, the largest cross-framework difference ($49.9\%$ LangGraph versus $24.7\%$ AutoGen).

For each execution, we identify the disclosing agent as the earliest to access the protected value via a vault channel, recover realized inter-agent edges ($C_2$) from the trace, and compute minimum graph distance from that agent to each topology-introduced recipient. We partition executions as \emph{direct-access} if at least one such recipient is one hop away, or \emph{relay-required} if every one is at least two hops away, capturing whether reaching them requires intermediate agents. Executions where the value never enters a vault channel have no defined disclosing agent or distance and are excluded from this analysis (retained in cell-level results): LangGraph has 62 such executions, leaving 423 with a defined disclosing agent, while AutoGen has none, leaving all 485. We report these denominators explicitly since the hop analysis is conditional on the value entering execution.

The decomposition shows substantially different patterns (Table~\ref{tab:relay-hop}): AutoGen exposure is $34.0\%$ for
direct-access executions versus only $2.8\%$ when relay-required, while LangGraph's corresponding rates are $55.9\%$ and $60.3\%$, so relay hops attenuate strongly in AutoGen but not LangGraph. The $95\%$ CI for AutoGen relay-required is $[1.1,7.0]\%$ versus
$[51.6,68.4]\%$ for LangGraph, widely separated intervals indicating a substantial conditional-exposure difference; LangGraph's direct-access and relay-required estimates are similar ($55.9\%$ versus $60.3\%$), showing no hop-based attenuation there. This refines the cell-level result: exposure depends not only on whether topology introduces additional recipients, but whether the value persists along the realized path to them.

\begin{table}[t]
\centering
\small
\caption{\small Realized-path decomposition of new-agent disclosure in Hier-7, measured from each framework's $C_2$ execution traces. Rates are reported with 95\% Wilson confidence intervals (CI). The analysis is conditional on a defined disclosing agent (423 LangGraph executions, 485 AutoGen executions).}
\vspace{-0.4cm}
\label{tab:relay-hop}
\begin{tabular}{@{}lll@{}}
\toprule
Hier-7 executions & AutoGen & LangGraph \\
\midrule
Direct-access   & 34.0\% [29.2,39.2], $n{=}341$ & 55.9\% [50.2,61.4], $n{=}297$ \\
Relay-required  & 2.8\% [1.1,7.0], $n{=}144$    & 60.3\% [51.6,68.4], $n{=}126$ \\
\midrule
Observed pattern & strong attenuation & no comparable attenuation \\
\bottomrule
\end{tabular}
\end{table}

\subsubsection{Evidence for a Relay-Semantic Explanation.}
\label{subsubsec:fw-mechanism}
The traces are consistent with different relay semantics across frameworks. In AutoGen, relay-required propagation involves successive inter-agent exchanges; the value must be retained and reproduced across multiple messages to reach a downstream recipient, and the lower relay-required exposure is consistent with propagation failures across these exchanges. LangGraph instead maintains shared
state along the relevant path, keeping earlier information available to downstream agents, consistent with showing no exposure reduction when recipients require additional relay hops.

A separate trace-level measure supports this: AutoGen relay failures occur in multi-hop CW-7 and Hier-7 at $12.2\%$ and $8.5\%$, while P2P cells, requiring no relay, show $0\%$ (Table~\ref{tab:relay-failure}). Relay failure is distinct from leakage: it records failure to propagate through a required relay operation, while leakage records whether the value is ultimately disclosed. These observations are consistent with a relay-semantic explanation for the AutoGen--LangGraph difference, though they do not establish message reproduction or shared-state execution as the sole cause; scheduling, state management, and execution control may also contribute.

\begin{table}[t]
\centering
\small
\caption{\small AutoGen relay-failure rate by topology cell. The P2P cells require
no relay.}
\vspace{-0.4cm}
\label{tab:relay-failure}
\begin{tabular}{@{}lr@{}}
\toprule
Topology cell & Relay-failure rate \\
\midrule
Hier-7 & 8.5\%  \\
CW-7   & 12.2\% \\
\midrule
P2P-4 / P2P-7 / P2P-10 & 0.0\% \\
\bottomrule
\end{tabular}
\end{table}

\subsubsection{Implication for Topology-Aware Privacy Enforcement.}
\label{subsubsec:fw-implication}
This analysis distinguishes \emph{nominal topology} from \emph{realized information flow}: the specified graph determines possible paths, but observed exposure also depends on how an orchestrator realizes and preserves information along them, so agent count, edge count, and topology class alone cannot characterize exposure. A sanitizer should therefore not infer authorization from a
protected value's presence or nominal topology alone; the decision should incorporate the intended recipient and observable communication context, an authorization decision being a fusion of value, realized recipient, and context. Therefore, we develop RECONS, operationalizing this by conditioning sanitization on recipient and observable execution context rather than a fixed transformation per detected value. The cross-framework results motivate the recipient- and context-conditioned enforcement evaluated in \S\ref{sec:sanitizer}.

\section{RECONS: Recipient-Conditioned Sanitization}\label{sec:sanitizer}

The evaluated multi-agent LLM workflows exhibit predominantly \emph{direct and propagative} leakage: a value disclosed by one agent can propagate through channels and execution steps to an agent unauthorized to receive it (\S\ref{sec:leakage-synthesis}). We did not observe successful reconstructive or behavioral side-channel leakage, which does not establish absence in general but identifies disclosure detection and propagation control as the primary
mitigation targets this evaluation supports.

\begin{figure*}
    \centering
    \small
    \includegraphics[width=0.8\linewidth]{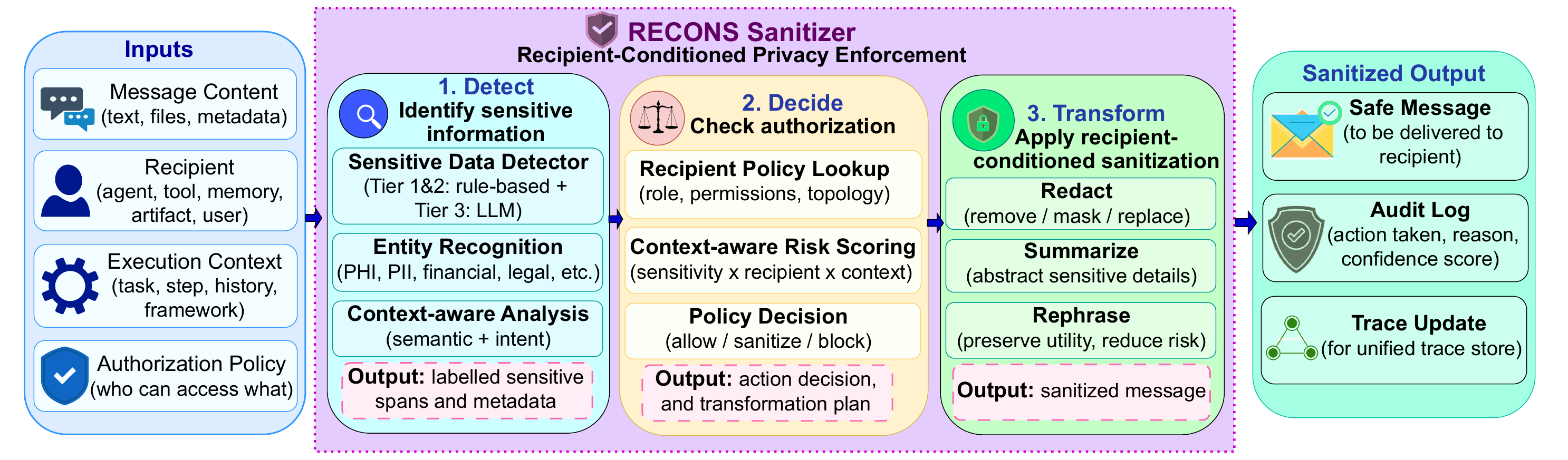}
    \vspace{-0.4cm}
    \caption{\small RECONS recipient-conditioned privacy enforcement. The sensitive information detector identifies what may be sensitive in the message. Policy decision engine determines what to do based on who can receive it and in what context. The sanitization module modifies the message according to the decision. Trace \& audit logger provides transparency and supports post-hoc auditing.}
    \label{fig:RECONS}
\end{figure*}

RECONS implements these as a three-stage pipeline, \textsc{Detect} $\rightarrow$ \textsc{Decide} $\rightarrow$ \textsc{Transform}, as shown in~\autoref{fig:RECONS}. Detection identifies candidate values and assigns a type and confidence score, decision determines whether the detected value is authorized for the intended recipient under observed context, and transformation applies a type-appropriate operation when disclosure is unauthorized, separating what information is present, whether it may be disclosed, and how unauthorized information should be transformed.

\subsection{Positioning}\label{subsec:sanitizer-positioning}
Existing mechanisms operate at the user--LLM boundary or apply recipient-independent transformations. AgentLeak's blanket-redaction interceptor removes detected content uniformly, costing $4.7$ points of task success; Presidio and commercial DLP similarly detect predefined types and apply fixed transformations; Pr$\in\in$mpt~\cite{Preempt} provides token-level FPE and metric-LDP but targets single-agent, single-prompt settings without modeling recipient-dependent disclosure across agents.

RECONS instead targets values authorized for one recipient but not another as they propagate through a workflow, centered on a \emph{recipient-aware authorization layer} extending a field-level allowed-set policy with the actual recipient when observable, coupled with a \emph{type-aware transformation layer}: FPE for format-preserving needs, pseudonymization for non-numeric
value-dependent attributes, mLDP where numeric perturbation is acceptable, contextual rewriting for sensitive free text, and redaction as fallback. RECONS operates as an \emph{inline interceptor}, evaluating each agent-authored event before it reaches a downstream agent, tool, memory component, or recipient, distinguishing it from post-hoc sanitization and letting transformed content participate in subsequent steps, including legitimate returns to an originating
agent.

\subsection{Architecture}\label{subsec:sanitizer-architecture}

RECONS processes events on $C_1$--$C_7$ in three stages:

\begin{itemize}[leftmargin=*]
\item \textbf{Detection.} Identifies candidate values, assigns a
type, and records confidence.

\item \textbf{Decision.} The authorization policy
$\mathcal{A}(\breve{a}, \varsigma, c, r, \mathit{pos})$ determines whether value $\breve{a}$ may be disclosed on channel $c$ to recipient $r$ at topology position $\mathrm{pos}$, passing authorized disclosures unchanged and forwarding unauthorized ones to transformation.

\item \textbf{Transformation.} Selects an operation by value type and
required utility: FPE for format-sensitive values, mLDP for numeric values where approximate utility suffices, pseudonymization for the corpus's non-numeric value-dependent attributes, contextual rewriting for free text, and redaction when no suitable transformation applies.
\end{itemize}

The three types are defined by the transformation they require: \emph{format-sensitive} values need representation preserved over content; \emph{value-sensitive} values contribute numerically or semantically to downstream computation and should be preserved approximately; \emph{context-sensitive} values derive sensitivity from meaning rather than surface form and require rewriting or removal. This typing interfaces detection with transformation
(\S\ref{sec:decision-layer}--\ref{sec:transformation-layer}, evaluated
in \S\ref{subsec:eval-main}).

\subsection{Detection Layer}\label{sec:detection-layer}

The detector must identify protected values across heterogeneous agent-authored channels while providing information sufficient for authorization and transformation. Since the corpus contains both structured values and contextually sensitive free text, a single mechanism is insufficient, so RECONS uses a three-tier detector progressing from cheap pattern-based detection to semantic analysis.

\subsubsection{Design.}\label{subsec:detection-design}

The detector comprises three tiers:

\begin{itemize}[leftmargin=*]
\item \textbf{Tier 1: rule and NER detection.} Pattern matching and named-entity recognition for structured values, SSNs, account and card numbers, phone numbers, identifiers, amounts, and dates, with checksum validation and domain-specific recognizers where generic rules fall short.

\item \textbf{Tier 2: zero-shot NER.} GLiNER~\cite{GLiNER} extends coverage to entities lacking reliable patterns or conventional rules, replacing a Tier-1 detection only when its confidence exceeds Tier-1's, otherwise retaining the higher-confidence Tier-1 result.

\item \textbf{Tier 3: contextual LLM detection.} A calibrated judge identifies information expressed semantically rather than by pattern or entity type, a pending lawsuit, a strategic decision, a health condition in prose. It operates on free-text channels $C_1$, $C_2$, $C_6$ and the semantic content of $C_5$ and $C_7$, using the pinned \textsc{Mistral-Small-24B} judge at threshold $\zeta=0.50$ with a fixed prompt hash calibrated on held-out labels; each detection yields a candidate $(\text{span},\text{value},\text{value\_type},\text{channel})$, where $\text{value\_type}\in\{\text{format\_dependent}, \text{value\_dependent},\text{contextual}\}$.
\end{itemize}

\noindent Each detection retains its source tier and confidence, treated as evidence rather than ground truth downstream, letting the decision layer account for reliability differences across channels and value types.

\subsubsection{Validation Protocol.}\label{subsec:detection-validation}
We validate on a development split, reserving held-out scenarios exclusively for end-to-end evaluation (\S\ref{subsec:eval-protocol}): 80 scenarios ($16.5\%$) development, 405 held-out. Across both $\tau$ and $\tau'$ executions, the detector processes $12{,}606$ agent-authored instances with known ground truth. Recall is computed over positive instances stratified by value type, domain, channel, and attribute; precision is estimated by manual annotation of sampled
detections. Protected attributes are enumerated before evaluation and assigned a value type, with domain-specific recognizers added where generic detection is insufficient.

\subsubsection{Results: Detection Varies by Value Type.}\label{subsec:detection-results}
Pre-tuning detection varies substantially by value type (Table~\ref{tab:detection-valuetype}): among $3{,}262$ positive instances, recall is $85.0\%$ for value-sensitive, $65.1\%$ for format-sensitive, and $36.0\%$ for context-sensitive values, with structured identifiers like \texttt{ssn}, \texttt{email}, \texttt{insurance\_id}, and \texttt{name} detected more reliably than weakly patterned numeric values and contextual attributes. This exposes a surface-level detection limitation: contextual attributes like a chief complaint or internal note can be privacy-sensitive without a distinctive lexical, entity, or formatting signal, making a semantic tier necessary rather than merely an enhancement.

\begin{table}[t]
\centering
\small
\caption{\small Pre-tuning detection recall by value type over 3{,}262 positive
instances.}
\vspace{-0.4cm}
\label{tab:detection-valuetype}
\begin{tabular}{lrr}
\toprule
Value type & Recall & Caught / total \\
\midrule
Value-sensitive    & 85.0\% & 731 / 860 \\
Format-sensitive   & 65.1\% & 1023 / 1571 \\
Context-sensitive  & 36.0\% & 299 / 831 \\
\midrule
Overall            & 62.9\% & 2053 / 3262 \\
\bottomrule
\end{tabular}
\end{table}

\noindent\textbf{LLM tier.}
Since rule and NER tiers achieve only $36.0\%$ recall on context-sensitive values, we evaluate the Tier-3 judge on 60 development instances, 40 positive (stratified between cases detected and missed by preceding tiers) and 20 negative controls: the judge identifies 39 of 40 positives with no false positives on controls, demonstrating semantic-tier utility for these sampled
cases without characterizing population-level precision or recall. The judge's operating point is inherited fr

\subsubsection{Channel Dependence.}\label{subsec:detection-channel}
Detection performance also depends on channel representation. Comparing format-sensitive values across channels to reduce value-type confounding (Table~\ref{tab:detection-channel}), recall is substantially lower on structured channels, $C_3$ tool arguments ($39.2\%$) and $C_5$ memory writes ($40.5\%$), than on free-text channels like $C_2$ ($85.3\%$) and $C_6$ ($83.7\%$). Since low detection confidence does not imply low privacy risk, and the weakest-detection channels are also where substantial disclosure occurs, RECONS treats confidence as channel-conditioned evidence rather than a direct risk proxy, using conservative fallback where detection and recipient resolution are insufficient.

\begin{table}[t]
\centering
\small
\caption{\small Pre-tuning recall for format-sensitive values across channels,
isolating channel representation from value-type composition.}
\vspace{-0.4cm}
\label{tab:detection-channel}
\begin{tabular}{llr}
\toprule
Channel & Kind & Recall \\
\midrule
$C_2$ (inter-agent)   & free-text  & 85.3\% \\
$C_6$ (logs)          & free-text  & 83.7\% \\
$C_7$ (artifacts)     & mixed      & 79.3\% \\
$C_1$ (final output)  & free-text  & 73.7\% \\
$C_4$ (tool outputs)  & mixed      & 71.6\% \\
$C_5$ (memory writes) & structured & \textbf{40.5\%} \\
$C_3$ (tool arguments) & structured & \textbf{39.2\%} \\
\bottomrule
\end{tabular}
\end{table}

\subsubsection{Precision and Tuning.}\label{subsec:detection-tuning}
An automatic span-overlap proxy estimates precision at $15.7\%$, but treats detections of untracked sensitive fields as false positives; manual inspection of 40 sampled non-target detections found $35\%$ actually contained sensitive content belonging to other fields, with remaining errors mostly field labels misclassified as values and agent identifiers like ``A1'' and ``A2'' misclassified as PII. A tuning pass on the development split fixed these failure modes without reducing recall anywhere: overall recall rose from $62.9\%$ to $66.7\%$, format-sensitive recall from $65.1\%$ to $73.0\%$, with
the largest gain in local phone numbers, $19.6\%$ to $99.3\%$, after adding a format-specific recognizer.

\subsubsection{Carried-Forward Limitations}\label{subsec:detection-limitations}
Three limitations carry forward to the decision layer: structured channels have lower recall than free-text channels for the same value type; agent identifiers and canary-like tokens can be textually ambiguous, creating corpus-specific precision-recall trade-offs; and contextual coverage depends on, and inherits the operating characteristics of, the calibrated Tier-3 judge. These motivate a decision policy operating under incomplete, uncertain observations rather than treating detector output as a complete representation of privacy risk.

\subsection{Decision Layer}\label{sec:decision-layer}

The decision layer is RECONS's policy component. Given a detected value, its type and confidence, channel, recipient, and topology position, it selects an action $\mathcal{A}(\breve{a}, \varsigma, c, r, \mathit{pos}) \in \{\textsc{allow},\textsc{fpe},m\textsc{ldp},\textsc{rewrite},\textsc{redact}\}$. It first determines whether disclosure is authorized, then selects a value-type-compatible operation when protection is required. The transformation layer (\S\ref{sec:transformation-layer}) executes the selected operation. This design targets propagative leakage by sanitizing a value reaching an unauthorized recipient while preserving authorized disclosures.

\subsubsection{Authorization Source.}\label{subsec:decision-authorization}

RECONS reuses the field-level release policy underlying the topology leakage metric (\S\ref{sec:rq1}): each value specifies its authorized agents, with authorization reassigned across topology cells so richer-topology-introduced agents are unauthorized by construction, the same ground truth used by
\texttt{new\_agent\_disclosed\_rate}. Recipient resolution applies only where an explicit inter-agent boundary exists in the corpus: \texttt{event\_recipient} identifies recipients for inter-agent messages ($C_2$), and memory reads ($C_5$-\texttt{memory\_read}), with no recipient defined for $C_1$, $C_3$, $C_4$, $C_5$-\texttt{memory\_write}, $C_6$, or $C_7$. So, recipient-aware enforcement operates only on $C_2$ and $C_5$ reads,
the rest using the field-level policy. This also explains why so few authorized cases exist for assessing over-sanitization: the policy authorizes relatively few fields per recipient, leaving only 17 should-allow cases in the evaluated subset (\autoref{tab:decision-confusion}).

\subsubsection{Static and Dynamic Policies.}\label{subsec:decision-policies}

We compare two policies differing only in recipient awareness. \emph{Static-}$\mathcal{A}$ applies the field-level allowed set
regardless of actual recipient, while \emph{Dynamic-}$\mathcal{A}$
adds recipient-specific authorization via three rules.

\begin{itemize}[leftmargin=*]
\item \textbf{Recipient authorization.} Allows a value if the resolved recipient is authorized; otherwise, sanitizing by value type, FPE for format-dependent, pseudonymization or mLDP for value-dependent, rewriting for contextual, redaction as fallback.

\item \textbf{Confidence-conditioned protection.} Defaults uncertain detections toward protection where a recipient is resolvable. This rule is inapplicable to $C_3$ because no recipient is defined for that channel and of limited use for $C_5$ reads in this split.

\item \textbf{Topology-position extension.} Sanitizes values graph-reachable from an authorized immediate recipient to an unauthorized agent. This extension is evaluated separately and excluded from the core policy.
\end{itemize}

\subsubsection{Decision-Layer Validation}\label{subsec:decision-validation}
We evaluate decision correctness on the development split independently of transformed outputs. Among $4{,}834$ protected-value occurrences, the detector identifies $3{,}400$ ($70.3\%$). Because recipient-aware decisions require a resolvable recipient, analysis is restricted to $2{,}109$ occurrences both
detected and resolvable on $C_2$ or $C_5$ reads. Thus, the results in this section are conditional on successful detection and recipient resolution. End-to-end performance is evaluated separately in \S\ref{subsec:eval-main}.

We distinguish \emph{under-sanitization}---a required-protection value
allowed, constituting a privacy failure---from \emph{over-sanitization}, an
authorized value unnecessarily sanitized, incurring a potential utility cost.

\textit{Recipient-aware authorization removes topology-induced under-sanitization.} Among $2{,}092$ occurrences requiring sanitization, Static-$\mathcal{A}$ allows 50 ($2.39\%$) to pass, all when a field-level-permitted value reaches an unauthorized topology-introduced agent, while Dynamic-$\mathcal{A}$ allows none (\autoref{tab:decision-confusion}), eliminating these by conditioning authorization on the resolved recipient. Neither policy over-sanitizes the 17 should-allow cases. However, given the small sample, this result indicates no observed over-sanitization rather than a utility advantage.

\begin{table}[t]
\centering
\small
\caption{\small Decision correctness on the detected, recipient-resolvable
development subset ($n=2{,}109$). Dynamic-$\mathcal{A}$ eliminates the
50 topology-induced under-sanitization cases observed under
Static-$\mathcal{A}$.}
\vspace{-0.4cm}
\label{tab:decision-confusion}
\begin{tabular}{lrr}
\toprule
& Static-$\mathcal{A}$ & Dynamic-$\mathcal{A}$ \\
\midrule
Under-sanitization & 2.39\% & \textbf{0.00\%} \\
\quad should-sanitize, allowed ($/2{,}092$) & 50 & \textbf{0} \\ \hline
Over-sanitization & 0.00\% & 0.00\% \\
\quad should-allow, sanitized ($/17$) & 0 & 0 \\
\bottomrule
\end{tabular}
\end{table}

\textit{Operation selection follows value type.}
The 2{,}092 sanitization decisions comprise 1{,}260 FPE, 480
mLDP/pseudonymization, and 352 rewriting operations. Each decision
selects the operation associated with its assigned value type, with no observed type-operation mismatch.

\subsubsection{Topology-Position Extension}\label{subsec:decision-pos}
\subparagraph{}
The topology-position extension tests whether authorization of the
immediate recipient is sufficient in richer communication graphs. Of
the 17 cases allowed by the core Dynamic-$\mathcal{A}$ policy,
unrestricted graph reachability reclassifies 14 as requiring
sanitization, corresponding to an 82\% over-sanitization rate among
these authorized cases. This result illustrates a limitation of unrestricted potential reachability. An authorized recipient may have a graph path to an unauthorized agent even when the value is never forwarded along that path. Therefore, applying reachability alone would impose unnecessary sanitization. The extension is consequently excluded from the core policy. A more targeted approach would condition topology analysis on observed forwarding or other evidence that the value is actually being propagated.

\subsection{Transformation Layer}\label{sec:transformation-layer}

The transformation layer executes the decision layer's selected action (\S\ref{sec:decision-layer}), passing content through unchanged when the decision is \textsc{allow}, without independently re-evaluating authorization; it applies operation-specific verification checks. We verify the mechanisms on a fixed 103-occurrence development sample from 12 scenario-topology-cell pairs, each re-evaluated using the same pinned Tier-3 detector used throughout, so results characterize mechanism behavior under the study's instrument rather
than an independent non-disclosure guarantee. End-to-end leakage is evaluated separately on the full held-out set in
\S\ref{subsec:eval-main}.

\subsubsection{Transformation Operations.}\label{subsec:transform-operations}

\textit{FPE for format-dependent values.} FPE transforms format-dependent values, identifiers, account numbers, dates of birth, and canary tokens, using FF3-1. The transformation is reversible with the encryption key for authorized downstream recovery. Development verification found a boundary case: short-domain attributes like the three-digit \texttt{credit\_score} fail FF3-1's minimum domain-size requirement, and an initial implementation padding and truncating ciphertext to restore original length broke invertibility. We removed truncation and retain the full padded ciphertext instead, treating these as representation-length exceptions rather than strict length-preserving
FPE. Evaluated FPE cases preserve the configured character set and remain exactly reversible.

\paragraph{Pseudonymization and mLDP for value-dependent values.} The
corpus's naturally occurring value-dependent attributes, \texttt{name}, \texttt{address}, \texttt{patient\_name}, are non-numeric. So, this path uses deterministic keyed pseudonymization, the same value mapping to the same pseudonym under a fixed key. No collisions were observed, though this is not a formal guarantee. A separate numeric mLDP path uses the Laplace mechanism with
$\varepsilon=1.0$ under the $\ell_1$ metric, verified synthetically since no corpus attribute exercises it. Under the specified sensitivity and metric assumptions, the Laplace mechanism provides the corresponding $\varepsilon$-DP guarantee. The privacy parameter is configurable to support deployment-specific privacy--utility trade-offs.

\paragraph{Verified rewriting for contextual values.} Contextual values are rewritten under a frozen prompt, each candidate checked by the pinned detector before acceptance, rejected and retried up to three times if at or above threshold before falling back to redaction. Development found two failure modes. First, an early prompt caused the model to reproduce a protected field's attribute name as filler, fixed by treating field names as non-output identifiers. Second, the detector can classify broader paraphrases as disclosures. For example, replacing \emph{pending harassment lawsuit} with \emph{a pending legal matter} scores $0.8\geq\zeta$. So, retries treat the detector as a conservative gate rather than assuming natural paraphrase is safe.
Since the rewriting model and judge are not independent, verification establishes only that outputs satisfy the pinned detector criterion, not independent semantic non-disclosure; the full evaluation also uses an exact-match backstop, accepting a rewrite only if it passes semantic verification and contains no literal protected value. This was introduced after full-scale evaluation found that literal echoes semantic verification alone admitted (\S\ref{subsec:eval-main}).

\paragraph{Redaction and fallback.} Redaction serves as fallback when a transformation cannot be safely applied, FPE failing on format, pseudonymization or mLDP failing to produce a valid result, or rewriting exhausting its attempts. Each fallback's source is recorded to distinguish transformation-specific failures from ordinary redaction.

\subsubsection{Mechanism Verification.}\label{subsec:transform-validation}
On the development sample (\autoref{tab:transform-summary}), all 103 transformed occurrences are classified as non-disclosing post-transformation ($100\%$ detector-based $P$-removal), specific to the detector rather than proof of semantic non-disclosure. The sample comprises 18 FPE, 50 mLDP/pseudonymization, and 35 rewriting decisions; of the latter, 27 ($77.1\%$) pass on the first attempt and 8 ($22.9\%$) exhaust all three and fall back to redaction, with no second- or third-attempt successes. No FPE or pseudonymization fallbacks in this sample, and no FPE format-recognition fallback across the full $2{,}595$-occurrence FPE population. Successful rewrites score $0.8$ to $0.95$ on a separate meaning-preservation evaluator, applied only to successful rewrites here and reported as supplementary characterization rather than an end-to-end utility estimate.

\begin{table}[t]
\centering
\footnotesize
\caption{\small Transformation-layer verification on the development sample
($n=103$). FPE checks combine 18 corpus-occurring and 27 synthetic
cases; other rows are scoped to the indicated sample or corpus
population.}
\vspace{-0.4cm}
\tabcolsep = 0.08cm
\label{tab:transform-summary}
\begin{tabular}{lr}
\toprule
Metric & Result \\
\midrule
Decided action: FPE / mLDP / rewrite & 18 / 50 / 35 \\
Rewrite outcome: first-attempt clear / redaction fallback & 27 (77.1\%) / 8 (22.9\%) \\
Detector-based $P$-removal & 103/103 (100\%) \\
\midrule
FPE round-trip: corpus / synthetic & 18/18 + 27/27 \\
FPE character-set preservation & 45/45 \\
FPE length preservation & 41/45 \\
Wrong-key recovery differs from original & 27/27 \\
FPE format-recognition fallback (full corpus, $n=2{,}595$) & 0.0\% \\
\midrule
Pseudonymization determinism / $P$-removal & 50/50 \\
mLDP (synthetic): $P(|\text{error}|\leq 10\text{k})$ & 1.000 \\
mLDP DP-inequality diagnostic & 23/24 \\
\midrule
\textsc{allow} passthrough (sampled) & 10/10 \\
\bottomrule
\end{tabular}
\end{table}

\subsubsection{Evaluation Protocol.}\label{subsec:eval-protocol}
All sanitizer components, thresholds, prompts, and policies are fixed using the 80-scenario development split. Evaluation uses all 405 held-out scenarios across six topology cells, yielding $2{,}430$ scenario-cell runs and $34{,}884$ protected-value occurrences. Since 105 scenarios share no common horizon with the rest of the corpus, each is evaluated at its canonical horizon where available, otherwise its shortest covered horizon, with no parameter re-tuned under this assignment. The implications of this horizon heterogeneity are
discussed in \S\ref{subsec:eval-limits}. Leakage is measured using the same pinned three-tier detector throughout, so pre- and post-sanitization measurements share the same evaluation instrument.

\noindent\textbf{Authorization-anchored metrics.}
We report two authorization-conditioned metrics: \emph{residual leakage}, the fraction of should-sanitize occurrences ($n=34{,}581$) still accessible to an unauthorized recipient after sanitization, and \emph{correct-allow}, the fraction of should-allow occurrences ($n=303$) still accessible to their authorized recipient. Lower residual leakage and higher correct-allow are preferable. Correct-allow measures authorized-disclosure retention, not general utility, anchored to authorization ground truth rather than task performance or content quality. We therefore separately report a content-level meaning-preservation proxy on a stratified 120-occurrence sample, with 40 per value type.

\noindent\textbf{Baselines.}
All conditions use identical held-out executions: \emph{None} performs no sanitization, \emph{blanket redaction}~\cite{AgentLeak} redacts every detected value regardless of recipient, \emph{Presidio}~\cite{Presidio} applies fixed-infoType DLP detection followed by uniform redaction, and \emph{RECONS} applies the proposed recipient-aware policy and type-aware transformation.  Presidio serves as a reproducible open baseline for recipient-blind, fixed-type DLP.

\subsection{RECONS Evaluation} \label{subsec:eval-main}
Table~\ref{tab:eval-main} summarizes the end-to-end results. RECONS reduces residual leakage from $96.7\%$ without sanitization to $29.25\%$, versus $28.3\%$ for blanket redaction, while retaining $94.1\%$ of authorized disclosures versus blanket redaction's $20.5\%$, approaching blanket redaction's leakage rate while
retaining substantially more authorized disclosures. Presidio performs worse on both, $35.5\%$ residual leakage and $34.7\%$ correct-allow. Among the 395 held-out scenarios with no legitimately disclosable field, RECONS records $28.0\%$ residual leakage versus blanket redaction's $28.4\%$, so the aggregate difference
concentrates in the 10 scenarios with legitimately disclosable fields, where RECONS preserves an authorized disclosure that blanket redaction suppresses by design.

\begin{table}[t]
\centering
\small
\caption{\small End-to-end results on the held-out set
(405 scenarios; 2{,}430 scenario--cell runs; 34{,}884
protected-value occurrences). Residual leakage is lower-is-better;
correct-allow is higher-is-better.}
\vspace{-0.4cm}
\label{tab:eval-main}
\begin{tabular}{lrr}
\toprule
Condition & Residual leakage $\downarrow$ & Correct-allow $\uparrow$ \\
\midrule
None                         & 96.7\%  & 0.0\%  \\
Presidio (DLP)               & 35.5\%  & 34.7\% \\
AgentLeak blanket redaction  & 28.3\%  & 20.5\% \\
\textbf{RECONS}            & \textbf{29.25\%} & \textbf{94.1\%} \\
\bottomrule
\end{tabular}
\end{table}

\subsubsection{Sources of Residual Leakage.}\label{subsec:eval-decomposition}
The decomposition localizes most residual cases to detection misses and recipient-unresolvable cases (Table~\ref{tab:eval-decomp}). Of RECONS's $10{,}116$ residual cases, $9{,}657$ ($95.5\%$) are undetected values that never reached the decision or transformation layers. A further $452$ ($4.5\%$) are detected values with no resolvable recipient on $C_1$, $C_3$, $C_4$, $C_6$, or $C_7$, using the field-level fallback, and seven cases ($0.07\%$ of residual cases, $0.02\%$ of all should-sanitize occurrences) are sanitizer-acted content the evaluation detector still flags, examined individually in\ \S\ref{subsec:eval-ablations} as post-transformation detector-positive rather than confirmed disclosures.

Detection coverage is thus the dominant source: $69.2\%$ of should-sanitize occurrences were detected on the held-out set versus $66.7\%$ recall on the development split (\S\ref{subsec:detection-tuning}), populations that need not match. While detection recall does not directly predict residual leakage given that channel, authorization, and transformation also matter, the sanitizer plainly cannot act on an undetected value. The 452 fallback cases are a separate limitation, since unresolvable recipients preclude recipient-specific authorization. Thus, detection coverage and recipient resolution address distinct sources of residual exposure.

\begin{table}[t]
\centering
\small
\caption{\small Decomposition of RECONS's 10{,}116 residual cases. Detection misses dominate; field-level fallback is the principal RECONS-specific source.}
\vspace{-0.4cm}
\label{tab:eval-decomp}
\begin{tabular}{lrr}
\toprule
Cause & Count & Share \\
\midrule
Detection miss (never flagged)  & 9{,}657  & 95.5\% \\
Field-level fallback (no resolvable recipient) & 452 & 4.5\%  \\
Post-transformation detector-positive & 7  & 0.07\% \\\midrule
Total & 10{,}116 & 100\%  \\
\bottomrule
\end{tabular}
\end{table}

The value-type breakdown agrees (Table~\ref{tab:eval-valuetype}): residual leakage is $17.6\%$ for value-dependent, $23.3\%$ for format-dependent, and $55.7\%$ for contextual values, the last consistent with contextual values' lower detection recall (\S\ref{sec:detection-layer}).

\begin{table}[t]
\centering
\small
\caption{\small RECONS residual leakage by protected-value type. Contextual values exhibit the highest residual exposure, consistent with weaker contextual detection.}
\vspace{-0.4cm}
\label{tab:eval-valuetype}
\begin{tabular}{lr}
\toprule
Value type & Residual leakage \\
\midrule
Value-dependent  & 17.6\% \\
Format-dependent & 23.3\% \\
Contextual       & 55.7\% \\
\bottomrule
\end{tabular}
\end{table}

\subsubsection{Channel-Level Disclosure.}\label{subsec:eval-channel}

\begin{figure}
    \centering
    \includegraphics[width=0.85\linewidth]{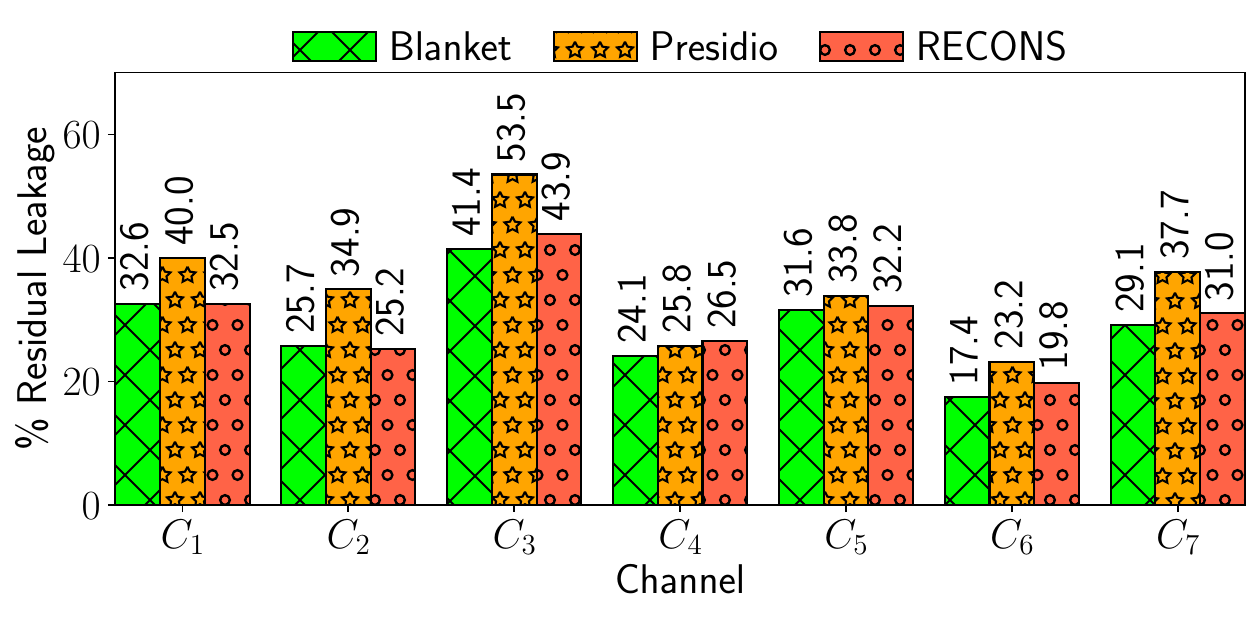}
    \vspace{-0.55cm}
    \caption{\small Residual leakage by observable channel. $C_3$ exhibits the highest residual leakage across the sanitizing conditions; RECONS trails blanket redaction on every channel except $C_1$ and $C_2$.}
    \label{fig:eval-channel}
\end{figure}

By channel (\autoref{fig:eval-channel}), $C_3$, structured tool-call arguments, shows the highest residual leakage across conditions, consistent with lower structured-channel recall (\S\ref{sec:detection-layer}) and field-level fallback for unresolvable $C_3$ recipients (\S\ref{sec:decision-layer}). RECONS matches or improves on blanket redaction only on $C_1$ and $C_2$ ($32.5\%$ vs.\ $32.6\%$; $25.2\%$ vs.\ $25.7\%$); elsewhere it trails: $C_3$ $43.9\%$ vs.\ $41.4\%$, $C_4$ $26.5\%$ vs.\ $24.1\%$, $C_5$ $32.2\%$ vs.\ $31.6\%$, $C_6$ $19.8\%$ vs.\ $17.4\%$, $C_7$ $31.0\%$ vs.\ $29.1\%$. The largest gap is on $C_3$, matching its
concentration of unresolvable cases.

% \begin{table}[t]
% \centering
% \small
% \caption{\small Residual leakage by observable channel. $C_3$ exhibits the highest residual leakage across the sanitizing conditions; RECONS trails blanket redaction on every channel except $C_1$ and $C_2$.}
% \vspace{-0.4cm}
% \label{tab:eval-channel}
% \begin{tabular}{lrrr}
% \toprule
% Channel & Blanket & Presidio & RECONS \\
% \midrule
% $C_1$ & 32.6\% & 40.0\% & 32.5\% \\
% $C_2$ & 25.7\% & 34.9\% & 25.2\% \\
% $C_3$ & 41.4\% & 53.5\% & 43.9\% \\
% $C_4$ & 24.1\% & 25.8\% & 26.5\% \\
% $C_5$ & 31.6\% & 33.8\% & 32.2\% \\
% $C_6$ & 17.4\% & 23.2\% & 19.8\% \\
% $C_7$ & 29.1\% & 37.7\% & 31.0\% \\
% \bottomrule
% \end{tabular}
% \end{table}

\subsubsection{Topology Gradient Under Sanitization.}\label{subsec:eval-gradient}

Next, we evaluate whether RECONS reduces disclosure to the topology-introduced agents identified in \S\ref{sec:rq1}. CW-2, with no topology-introduced agents, gives a $0\%$ reference point. Across the five richer topologies ($8{,}873$ new-agent-recipient occurrences), unsanitized disclosure is $97.6$--$98.0\%$ versus RECONS's $25.1$--$26.9\%$ (Table~\ref{tab:eval-gradient}). RECONS beats blanket redaction in four of five non-trivial cells and ties in Hier-7. Since topology-introduced agents are unauthorized under both, this does not isolate recipient-aware authorization's contribution, which instead shows in authorized-disclosure retention (Table~\ref{tab:eval-main}). Presidio leaves a higher residual floor, $34.2$--$35.8\%$, across these cells.

\begin{table}[t]
\centering
\tabcolsep = 0.1cm
\small
\caption{\small Disclosure to topology-introduced agents before and after sanitization.}
\vspace{-0.4cm}
\label{tab:eval-gradient}
\begin{tabular}{lrrrrrr}
\toprule
Condition & CW-2 & CW-7 & P2P-4 & P2P-7 & P2P-10 & Hier-7 \\
\midrule
None     & 0\% & 98.0\% & 97.9\% & 98.0\% & 97.6\% & 97.8\% \\
Presidio~\cite{Presidio} & 0\% & 34.2\% & 34.7\% & 35.1\% & 35.8\% & 35.0\% \\
Blanket Redaction~\cite{AgentLeak}  & 0\% & 26.7\% & 27.0\% & 25.5\% & 25.9\% & 26.5\% \\
RECONS & 0\% & 25.9\% & 26.9\% & 25.1\% & 25.4\% & 26.5\% \\
\bottomrule
\end{tabular}
\end{table}

\subsubsection{Ablation Analysis.}\label{subsec:eval-ablations}

\paragraph{\textbf{Dynamic versus static authorization.}} Replacing recipient-aware authorization with a static allow-list raises residual leakage from $29.25\%$ to $29.81\%$, $0.56$ points, while raising correct-allow from $94.1\%$ to $100\%$. The static policy preserves all legitimately allowed fields regardless of recipient. The two policies differ mainly on the 10 scenarios with
legitimately disclosable fields, limiting the effective sample size of this comparison. Within this scope, the result is consistent with recipient awareness sharpening field- versus recipient-level authorization.

\paragraph{\textbf{Transformation menu versus redaction.}}
Replacing the FPE, pseudonymization, and rewriting menu with uniform redaction under the same decisions changes residual leakage only marginally, from $29.25\%$ to $29.59\%$, with correct-allow unchanged at $94.1\%$. Because both conditions share authorization decisions, this result indicates that the transformation menu's role is preserving authorized information and utility rather than lowering leakage itself.

\paragraph{\textbf{Transformation verification.}}
At full scale, semantic-only rewrite verification admitted some literal copies of the protected value. So, we added an exact-match backstop requiring a rewrite to pass semantic checking and contain no literal occurrence, falling back to redaction otherwise. This cut post-transformation detector-positive cases from 67 to 7, an $89.6\%$ reduction. Of the seven remaining cases, six involve negated true values, such as ``no criminal history,'' where rewrites contain no literal occurrence but the judge still classifies them as disclosures, reflecting
reference-to-topic versus disclosure-of-value. The seventh reflects disagreement between two calls to the same judge on identical, literal-value-free content, consistent with batched-inference non-determinism at temperature zero. None reflects literal reintroduction bypassing the exact-match backstop, though this establishes only that under this instrument, not immunity to semantic or inferential leakage.

\subsubsection{Utility and Evaluation Limitations}\label{subsec:eval-limits}

Table~\ref{tab:eval-utility} reports the content-level meaning-preservation proxy. The proxy assigns higher scores to blanket redaction ($99.3\%$) than RECONS ($83.5\%$), with the largest gap for contextual values ($69.1\%$), possibly reflecting the proxy's sensitivity to surface-form change: FPE and rewriting alter textual representations, while redaction can leave syntactically coherent output despite removing information. RECONS scores higher for format-dependent ($91.6\%$) and value-dependent ($89.6\%$) values, where FPE and pseudonymization apply. Since the proxy does not directly measure structured-field usability or task-level utility, we do not read the $99.3\%$ versus $83.5\%$ gap as evidence of greater practical utility for blanket redaction; correct-allow remains the primary retention measure, being directly anchored to ground-truth authorization labels.

Overall, recipient-aware sanitization approaches blanket redaction's residual-leakage performance while substantially increasing authorized-disclosure retention. The dominant remaining limitation is detection coverage, particularly for contextual values, with recipient resolution a separate limitation of the current policy.

\begin{table}[t]
\centering
\small
\caption{\small Content-level meaning-preservation proxy by condition
(stratified sample, $n=120$; 40 per value type) and by value type for
RECONS. The proxy does not fully characterize structured-field
usability or task-level utility.}
\vspace{-0.4cm}
\label{tab:eval-utility}
\begin{tabular}{lr}
\toprule
Condition & Utility \\
\midrule
AgentLeak blanket redaction          & 99.3\% \\
Presidio (DLP)                       & 97.3\% \\
RECONS (redact-only ablation)      & 98.8\% \\
\textbf{RECONS}                    & \textbf{83.5\%} \\
\midrule
\multicolumn{2}{l}{\emph{RECONS, by value type:}} \\
\quad Format-dependent (FPE)         & 91.6\% \\
\quad Value-dependent (pseudonym.)   & 89.6\% \\
\quad Contextual (rewrite / redact)  & 69.1\% \\
\bottomrule
\end{tabular}
\end{table}

\section{Discussion and Limitations} \label{sec:discussion}

\subsection{Design Implications}\label{subsec:discussion-evidence}

Our findings suggest privacy protection in MALS cannot be designed from static architectural assumptions and must instead be enforced against the realized recipient and execution path at runtime. The evidence supports three bounded conclusions: topology changes protected-value recipient exposure; recipient-conditioned enforcement
improves the privacy--authorization trade-off relative to blanket redaction; and detection coverage is the principal remaining bottleneck, with recipient resolution a secondary boundary. The rest of this section develops each in turn with its bounding limitations.

 \subparagraph{\textbf{Topology as a security-relevant configuration.}}
Topology-driven exposure (\S\ref{sec:rq1}) is not an artifact a better detector or implementation would remove but a structural consequence of which agents a communication graph makes reachable. Surface leakage varies---whether a value is disclosed at all---by only $2.2$ points across six cells, while new-agent disclosure---whether that value reaches a recipient the topology introduces---rises from a structural $0\%$ to $52.2\%$ over the same cells. So topology belongs in a system's security configuration alongside access control, not only performance or scalability
planning. We claim no universal topology law: ranking among richer topologies is unstable across frameworks and models (\S\ref{sec:cross-framework}, \S\ref{subsec:model-generalization}), and no one topology, hierarchical delegation included, is asserted as intrinsically riskier. What generalizes is that richer topology creates exposure a simpler one structurally cannot, not its magnitude
or ordering.

\subparagraph{\textbf{Recipient Conditioning as the Key Design Principle.}}
RECONS's decision layer implements one conditional rule, $\text{unauthorized} \rightarrow \textsc{sanitize}, \;\text{authorized} \rightarrow \textsc{allow}$, the entire mechanism separating recipient-conditioned from recipient-blind enforcement applied to the same detected value. Its advantage is not dramatically lower leakage; RECONS's residual $29.25\%$ sits close to blanket redaction's $28.3\%$ since both discard topology-introduced disclosure identically. The advantage is selective enforcement, which retains $94.1\%$ of authorized disclosures against $20.5\%$ by permitting exactly what a recipient-blind policy cannot distinguish from a violation. Recipient conditioning should be judged on what it preserves, not how far it lowers an already detection-bounded rate. This comparison rests on a modest base, only $303$ should-allow occurrences, with the RECONS-versus-blanket difference concentrated in just $10$ of $405$ scenarios containing a legitimately disclosable
field, so the $94.1\%$ versus $20.5\%$ gap should be read as consistent with the underlying behavior rather than as a precise population estimate.

%consistent with, not a precise estimate of, the underlying behavior.

\subparagraph{\textbf{Detection as the Principal Bottleneck.}}
The RECONS pipeline's detect-decide-transform structure means transformation
can only apply to a value the detector already flagged: an undetected value never
reaches the decision layer and passes through unchanged regardless of transformation sophistication. This is architectural, not correlational: $95.5\%$ of end-to-end residual leakage (\S\ref{subsec:eval-decomposition}) traces to exactly this: disclosures never flagged, concentrated in contextual attributes with the weakest semantic recall, a ceiling any same-detector sanitizer, blanket redaction included, would share, since neither can act on what neither sees. Only improved detection, not decision or transformation refinement, recovers this loss.

\subparagraph{\textbf{Recipient Resolution as an Architectural Boundary.}}
Recipient-aware authorization presupposes a resolvable recipient. $C_2$ messages and $C_5$ reads satisfy this directly; $C_3$ tool-call arguments target a tool, not an agent, so $95.8\%$ of $C_3$ events are self-loops crossing no inter-agent boundary, with the remaining $4.2\%$ reaching another agent via an already-covered $C_5$
read. Hence, field-level fallback on $C_3$ is correct given the channel's structure, not a coverage gap. Extending recipient resolution to the small residual of genuinely tool-mediated disclosure would require richer provenance than call-and-response. Specifically, tracking which agent consumes a tool's output when not itself a logged $C_5$ read, and communication semantics that treat a tool's result as attributable to whichever agent next acts on it.  Absent that provenance, recipient resolution cannot be extended to this channel without inventing a recipient the trace does not evidence.

\subparagraph{\textbf{Horizon Heterogeneity.}}
Each scenario is evaluated at its shortest available horizon (\S\ref{subsec:corpus-construction}), unevenly across domains: 105 scenarios lack a horizon common to the rest, and even fully-covered domains skew, finance and healthcare almost entirely at $T_1$, legal predominantly at $T_3$ or longer. So, cross-domain comparisons here
are confounded with horizon length and should not be read as horizon-controlled domain estimates.

\subsection{Threat-Model Limitations}\label{subsec:discussion-threat-limits}
Our threat model, and consequently every result in this paper, explicitly inherits the following limitations:
\begin{itemize}[leftmargin=*]
    \item \textbf{Corpus and generalization scope}: experiments derive from an English-language, four-domain subset of one public benchmark~\cite{AgentLeak}, with one generator and framework as the primary configuration; cross-model and
    cross-framework checks cover a handful of additional systems each, showing findings are not artifacts of one system but not that they generalize beyond those tested.
    \item \textbf{Latent-state leakage}: a value influencing a model's later output through its own internal context or reasoning state rather than an explicit channel event. Hence, it cannot be detected, attributed to a recipient, or evaluated by a trace-based methodology like ours.
    
    \item \textbf{Stronger collusion}: unauthorized recipients pooling their individual observations.
    \item \textbf{Cross-session inference}: an adversary correlating evidence across independent executions rather than within one trace.
    \item \textbf{Unobserved side channels}: any channel beyond the seven instrumented ($C_1$--$C_7$).
    \item \textbf{Attacks outside the evaluated threat model}: reconstruction from indirect evidence and behavioral side-channels, tested directly with no confirmed instance found
    (\S\ref{subsec:reconstruction-results}, \S\ref{sec:behavioral-results}), though this does not establish absence under a more capable or adaptive adversary.
\end{itemize}

\section{Conclusion} \label{sec:conclusion}
We investigated privacy leakage in multi-agent LLM systems as a question of \emph{where} a protected value travels, not only whether it is disclosed. Across six topology cells, topology changes recipient exposure rather than overall disclosure incidence, rising from $0\%$ to $52.2\%$ while surface leakage moves only $2.2$ points, holding when replicated across additional models and frameworks. In
a separate experiment holding topology fixed, framework choice alone produced a further two-orders-of-magnitude spread, demonstrating that architecture, not just communication structure, drives exposure. Exhaustive testing across three reconstruction axes and a behavioral side-channel found no confirmed leakage beyond direct disclosure and its propagation, confirming this is where these systems actually
leak, not a scope chosen for convenience. We addressed this with RECONS, an inline \textsc{Detect} $\to$ \textsc{Decide} $\to$ \textsc{Transform} sanitizer conditioning enforcement on recipient rather than sensitivity alone, reducing residual disclosure to within one point of blanket redaction ($29.25\%$ versus $28.3\%$) while
retaining $4.6\times$ as many authorized disclosures ($94.1\%$ versus $20.5\%$), showing that suppressing unauthorized disclosure need not suppress what
policy already permits.

Detection, not recipient resolution, is the binding constraint: $95.5\%$ of residual cases are values no same-detector sanitizer could act on, a limitation any comparable system shares, while only $4.5\%$ trace to RECONS's own resolution step. Recipient-conditioned enforcement is therefore necessary but not sufficient; closing the gap
depends more on detection coverage, particularly for contextual attributes, than on refining the decision layer. More broadly, privacy controls for multi-agent LLM systems cannot be assessed or built from static topology, framework, or model assumptions; they must act on the recipient and communication path realized at
runtime.

\balance
\bibliographystyle{ACM-Reference-Format}
\bibliography{references, references_1}

\clearpage
\appendix
\section{Appendix}

\begin{table}[H]
\centering
\caption{\small Interaction archetypes used in trajectory construction.}
\vspace{-0.4cm}
\label{tab:archetype-defs}
\small
\begin{tabular}{@{}lp{5.6cm}@{}}
\toprule
Archetype & Working definition \\
\midrule
BASE (initial task) & The trajectory's opening turn, in which the acting agent is given the task and the protected value. \\
A (elaboration) & A turn adding detail or context to a prior statement, without being directly asked to confirm or reveal the protected value. \\
B (handoff) & A turn transferring the task, case, or context to another agent. \\
C (justification) & A turn explaining or defending a decision or recommendation. \\
D (error handling) & A turn responding to or correcting a mistake or inconsistency earlier in the trajectory. \\
E (closeout) & A turn concluding the trajectory or interaction, e.g.\ summarizing the outcome. \\
F (rapport) & A turn oriented toward social or relational exchange rather than task content. \\
G (confirmation probe) & A turn in which one agent asks another to confirm, deny, or restate a specific candidate value, directly soliciting disclosure if the candidate is correct. \\
\bottomrule
\end{tabular}
\end{table}

\begin{table}[H]
\centering
\small
\caption{\small \textsc{Gemini-3.7-Flash} Hier-7 disclosure rates by protected
attribute type. Structured identifiers exhibit lower observed disclosure
rates than contextual and narrative attributes.}
\vspace{-0.4cm}
\label{tab:gemini-attribute}
\begin{tabular}{lr}
\toprule
Attribute & Disclosure rate \\
\midrule
SSN / DOB / phone / email / account & 0.0\% \\
Salary / balance / credit score     & 0.0\% \\
Address                             & 0.9\% \\
Settlement amount                   & 3.7\% \\
\midrule
Confidential strategy               & 27.6\% \\
Internal notes                      & 36.4\% \\
Criminal history                    & 63.6\% \\
\bottomrule
\end{tabular}
\end{table}

\end{document}